\documentclass[twocolumn,trackchanges,tighten]{aastex63}
\usepackage{graphicx}
\usepackage{natbib}
\usepackage{url}
\usepackage{footnote}
\usepackage{blindtext, rotating}
\usepackage[T1]{fontenc}
\usepackage{amsmath,amssymb}
\usepackage{cleveref}
\usepackage{float} 
\usepackage[caption=false]{subfig}

\received{}
\revised{}
\accepted{}
\begin{document}
\title{ Ionized gas metallicity of the strong [O~\textsc{iii}]$\lambda$5007 emission-line compact galaxies in the LAMOST survey}
\correspondingauthor{A-Li Luo, Wei Zhang}
\email{* lal@nao.cas.cn, xtwfn@bao.ac.cn }
\author[0000-0002-5345-4175]{Siqi Liu}
\affiliation{CAS Key Laboratory of Optical Astronomy, National Astronomical Observatories, Beijing 100101, China}
\affiliation{School of Astronomy and Space Science, University of Chinese Academy of Sciences, Beijing 100049, China}
\author[0000-0001-7865-2648]{A-Li Luo$^{*}$}
\affiliation{CAS Key Laboratory of Optical Astronomy, National Astronomical Observatories, Beijing 100101, China}
\affiliation{School of Astronomy and Space Science, University of Chinese Academy of Sciences, Beijing 100049, China}
\affiliation{College of Computer and Information Management \& Institute for Astronomical Science, Dezhou University, Dezhou 253023, China}
\author[0000-0002-1783-957X]{Wei Zhang$^{*}$}
\affiliation{CAS Key Laboratory of Optical Astronomy, National Astronomical Observatories, Beijing 100101, China}\author[0000-0001-8011-8401]{Xiao Kong}
\affiliation{CAS Key Laboratory of Optical Astronomy, National Astronomical Observatories, Beijing 100101, China}
\author{Yan-Xia Zhang}
\affiliation{CAS Key Laboratory of Optical Astronomy, National Astronomical Observatories, Beijing 100101, China}
\affiliation{School of Astronomy and Space Science, University of Chinese Academy of Sciences, Beijing 100049, China}
\author[0000-0002-3073-5871]{Shi-Yin Shen}
\affiliation{Key Laboratory for Research in Galaxies and Cosmology, Shanghai Astronomical Observatory, Chinese Academy of Sciences, 80 Nandan Road, Shanghai,  200030, China}
\affiliation{Key Lab for Astrophysics, Shanghai, 200034, China}
\author{Yong-Heng Zhao}
\affiliation{CAS Key Laboratory of Optical Astronomy, National Astronomical Observatories, Beijing 100101, China}
\affiliation{School of Astronomy and Space Science, University of Chinese Academy of Sciences, Beijing 100049, China}

\begin{abstract}
This article reports a sample of 1830 strong [O~\textsc{iii}]$\lambda$5007 emission-line compact galaxies discovered with the LAMOST spectroscopic survey and the photometric catalog of SDSS.  
We newly identify 402 spectra of 346 strong [O~\textsc{iii}]$\lambda$5007 emission-line compact galaxies by finding compact isolated point sources.  
Combined with the samples in our previous work \citep{2022ApJ...927...57L}, this returns a sample of 1830 unique strong [O~\textsc{iii}]$\lambda$5007 emission-line compact galaxies with 2033 spectra of $z\le 0.53$.
For the sources with $2\sigma$ [O~\textsc{iii}]$\lambda$4363 detections, we calculate the gas-phase metallicity with the direct-$T_e$ method, and verify that the strong-line metallicity diagnostics calibrated with the direct-$T_e$ method also applies to this sample. 
The strong [O~\textsc{iii}]$\lambda$5007 emission-line compact galaxies fall below several $T_e$-calibrated mass-metallicity relations. 
The N/O measurements of the strong [O~\textsc{iii}]$\lambda$5007 emission-line compact galaxies mainly locate at a plateau at low metallicity, indicating the product of primary nucleosynthesis.
The Ne3O2 and O32 relation follows a tight linear relation with no redshift evolution.
The Ne3O2 anti-correlates with the stellar mass, and at fixed stellar mass the Ne3O2 increase with the redshift.
Eight sources with asymmetric [O~\textsc{iii}]$\lambda5007$ emission-line profiles have been identified, however with no [O~\textsc{iii}]$\lambda4363$ detection, which proves the rich metal content and complex ionized gas kinematics within the galaxies. 
Higher-resolution spectroscopy will be necessary to identify the ionized gas components in detail. 
\end{abstract}


\section{Introduction}
Strong [O~\textsc{iii}]$\lambda$5007 emission-line compact galaxies are well-known for their unique color, compactness, high star formation rate (SFR), low metallicity, and being in isolated environments \citep{2009MNRAS.399.1191C}.
According to the redshifts and correspondingly the change of color, strong [O~\textsc{iii}]$\lambda$5007 emission-line compact galaxies are also called Blueberry galaxies (at $z\le 0.05$), Green Pea galaxies ($0.112 \le z < 0.36$) and Purple Grape galaxies ($0.05< z < 0.112$ and $0.36\le z \le 0.53$). 

In our previous work \citep{2022ApJ...927...57L}, we selected 1694 spectra with 1547 unique objects from the dedicated and non-dedicated survey in the LAMOST spectra database \citep{1996ApOpt..35.5155W,2004ChJAA...4....1S,2012RAA....12.1243L}. 
We further analyze the star formation rate (SFR), the mass-metallicity relation (MZR), and the environment of these samples.
In this work, in joint with newly identified strong [O~\textsc{iii}]$\lambda$5007 emission-line compact galaxies, we compile a large sample of the strong [O~\textsc{iii}]$\lambda5007$ emission-line compact galaxies and discuss the metallicity of the ionized gas.
In the following discussion, these samples are called strong [O~\textsc{iii}]$\lambda$5007 emission-line compact galaxies, which consist of 2033 spectra of 1830 unique galaxies. 

The galaxies' gas-phase metallicity and ionization parameter provides rich information about their chemical evolution, among which gas-phase oxygen abundances are the best approach to measure the current metallicity.
Emission lines in the spectra of the star-forming galaxies are acquired conveniently across different redshift ranges and thus are potent indicators of chemical compositions.
The well-known scaling relation between the gas-phase metallicity and the stellar mass, quoted as the mass-metallicity relation (MZR, \citet{1979A&A....80..155L}), has been studied extensively \citep{2004ApJ...613..898T,2006ApJ...647..970L,2010MNRAS.408.2115M,2010A&A...521L..53L,2013ApJ...765..140A,2019A&ARv..27....3M,2020A&A...634A.107Y,2020MNRAS.491..944C,2022ApJS..262....3N}. 
The MZR brings crucial observational information on understanding the build-up of the galaxies over time.
Typically there are two methods to measure the metallicity: the direct-$T_e$ method and the strong-line method \citep{2013ApJ...765..140A}. 
The direct-$T_e$ method employs the ionized gas's electron temperature and thus the gas's metallicity, calculated from the flux ratio of the auroral line and the strong lines. 
It is effective because metals are the ionized gas's primary coolants. 
The auroral and strong lines originate from the second and excited states separately, where the electron temperature can be calculated from the relative ratios of these two populations.
The metallicity is strongly correlated with the electron temperature, where the low electron temperature corresponds to the high metallicity.
It is suggested by \citet{1979A&A....78..200A, 1979MNRAS.189...95P} to use some emission lines to calibrate the oxygen abundances, usually referred to as the ``strong-line method''.
The strong-line method does not measure the metallicity directly. 
Due to their sensitivity to metallicities, the strong lines are easier to measure when the auroral lines are too weak for high-metal sources.
However, there are systematic discrepancies \citep{2008ApJ...681.1183K,2012MNRAS.426.2630L} comparing the metallicity determined from the direct-$T_e$ method and the strong-line calibrations.
Systematical ``strong-line" calibrations have been performed \citep{2008A&A...488..463M,2015ApJ...813..126J,2017MNRAS.465.1384C,2018MNRAS.481.3520P,2020MNRAS.491..944C}, however, more tests at the low-metallicity regimes are needed.

An alternative chemical abundance diagnostic is nitrogen, probed by N/O, which compared with oxygen produced from Helium fusion, the formation is more complex. 
N/O is sensitive to the chemical evolution history of a galaxy because nitrogen is the product of a primary nucleosynthetic product (independent of metallicity) and a secondary nucleosynthetic product (dependent on the metallicity of the gas cloud) \citep{2022MNRAS.512.2867H}. 
Studies have shown at low metallicities, the N/O ratios stay at a plateau and do not vary with O/H; while towards higher metallicity as the galaxies evolve, the N/O begins to increase with O/H \citep{2005A&A...437..849S,2012MNRAS.424.2316P,2013ApJ...765..140A,2022MNRAS.512.2867H}. 
The samples in this work will provide valuable information about the N/O vs O/H relation at the low metallicity regime.

The [Ne \textsc{iii}]$\lambda$3869 lines serve as an additional crucial probe of the ionized gas.
Ne3O2 ([Ne {\sc iii}]$\lambda3869$/[O {\sc ii}]$\lambda\lambda 3727,3729$) was first initiated as the metallicity diagnostic \citep{2006A&A...459...85N, 2007A&A...475..409S}.
However, \citet{2007MNRAS.381..125P,2014ApJ...780..100L} addressed the reason why Ne3O2 anti-correlates with the metallicity is the change of the ionization parameter.
Compared with the commonly used O32 ([O {\sc iii}]$\lambda5007$/[O {\sc ii}]$\lambda\lambda 3727,3729$), Ne3O2 is less prone to dust extinction due to the proximity of the wavelength coverage and can be applied to sources of higher redshifts. 
The strong emission-line samples in this work provide ample samples to re-examine the correlation of these two ionization parameters.

We also look into the kinematics of the ionized gas which reflects the star formation history. 
There are detailed optical \citep{2010ApJ...715L.128A,2012ApJ...749..185A,2012ApJ...754L..22A,2019MNRAS.489.1787B,2020MNRAS.494.3541H} and UV \citep{2014ApJ...791L..19J,2015ApJ...809...19H,2016ApJ...820..130Y,2018MNRAS.474.4514I} spectroscopic studies of Green Pea galaxies, where the authors have identified double-peak or asymmetric profiles of the emission lines.  
\citet{2012ApJ...754L..22A} has identified the spatially offset H$\alpha$ emission line profiles in their 2D spectra. 
\citet{2019MNRAS.489.1787B} analyzed the kinematics of a single Green Pea galaxy with the Gemini Multi-Object Spectrograph and confirmed the existence of three components: a rotating disk, a turbulent mixing layer, and gas outflow.  
\citet{2020MNRAS.494.3541H} concluded the triple component might come from two starbursts by fitting three components from all the optical emissions lines from deep high-resolution spectra. 
This has revealed the presence of complex ionized gas structures within these galaxies.

The structure of the paper is summarized below.  
In Section \ref{sec:data}, we introduce the newly spectroscopically-confirmed strong [O~\textsc{iii}]$\lambda$5007 emission-line compact galaxies from LAMOST.
In Section \ref{sec:para}, we explain how to measure the stellar mass from multi-wavelength photometry, obtain the emission line flux and the corresponding error, and measure the velocity dispersion from the [O~\textsc{iii}]$\lambda$5007 emission line.
In Section \ref{sec:results}, we demonstrate the result of the metallicity from the direct-$T_e$ method and compare it with the strong-line calibrations.
In Section \ref{sec:discussions}, we discuss the mass-metallicity relation, the N/O vs O/H relation, the Ne3O2 vs O32 relation, the Ne3O2 vs stellar mass relation, and the asymmetric [O~\textsc{iii}]$\lambda$5007 profiles of eight sources.
Finally, we present a summary of the measurements, results, and discussions in Section \ref{sec:conclusions}.

In this work, we use Wilkinson Microwave Anisotropy Probe (WMAP) 9 cosmology \citep{2013ApJS..208...19H}($\rm H_0 = 69.3 ~km~ s^{-1}Mpc^{-1} $, $\rm \Omega_m = 0.287$ and flat Universe), AB magnitudes \citep{1983ApJ...266..713O},
a Chabrier initial mass function (IMF; \citet{2003PASP..115..763C}), and \citet{bruzual2003stellar} stellar population synthesis (SPS) models.

\section{Sample and Data}
\label{sec:data}
\subsection{LAMOST spectra}
\subsubsection{Sample selection}
Instead of the traditional color selection criteria as in \citet{2009MNRAS.399.1191C}, we select the strong [O~\textsc{iii}]$\lambda5007$ emission line galaxies, \textbf{ which are compact and isolated.}
The reasons for removing the strict color selection are two-fold: to include the emission line galaxies with a broader redshift range and to select the galaxies with less severe emission lines. 

We select the sources from SDSS \texttt{CasJob} with the following criteria, where ``?" represents each of the $ugriz$ SDSS bands:
\begin{itemize}
    \item \texttt{petroRad\_r<3, -- compact object}
    \item \texttt{specObjID=0, --no spectrum has been taken in SDSS}
    \item \texttt{clean=1, -- reliable photometry}
    \item \texttt{mode<3 , --primary or secondary object}
    \item \texttt{parentID=0, -- no parent (if the object is deblended) or bright detection (if the object has one parent)},
    \item \texttt{petroRadErr\_?>0, -- morphology is certain in these five bands,}
    \item \texttt{petroR50Err\_?>0}
    \item \texttt{petroR90Err\_?>0}
    \item \texttt{cModelMagErr\_g<0.2, cModelMagErr\_r<0.2, cModelMagErr\_i<0.2}, 
    \item \texttt{17<cModelMag\_r<22}.
\end{itemize}
This returns 594863 sources.

\begin{figure}[!h]
  \centering
  \includegraphics[width=\hsize]{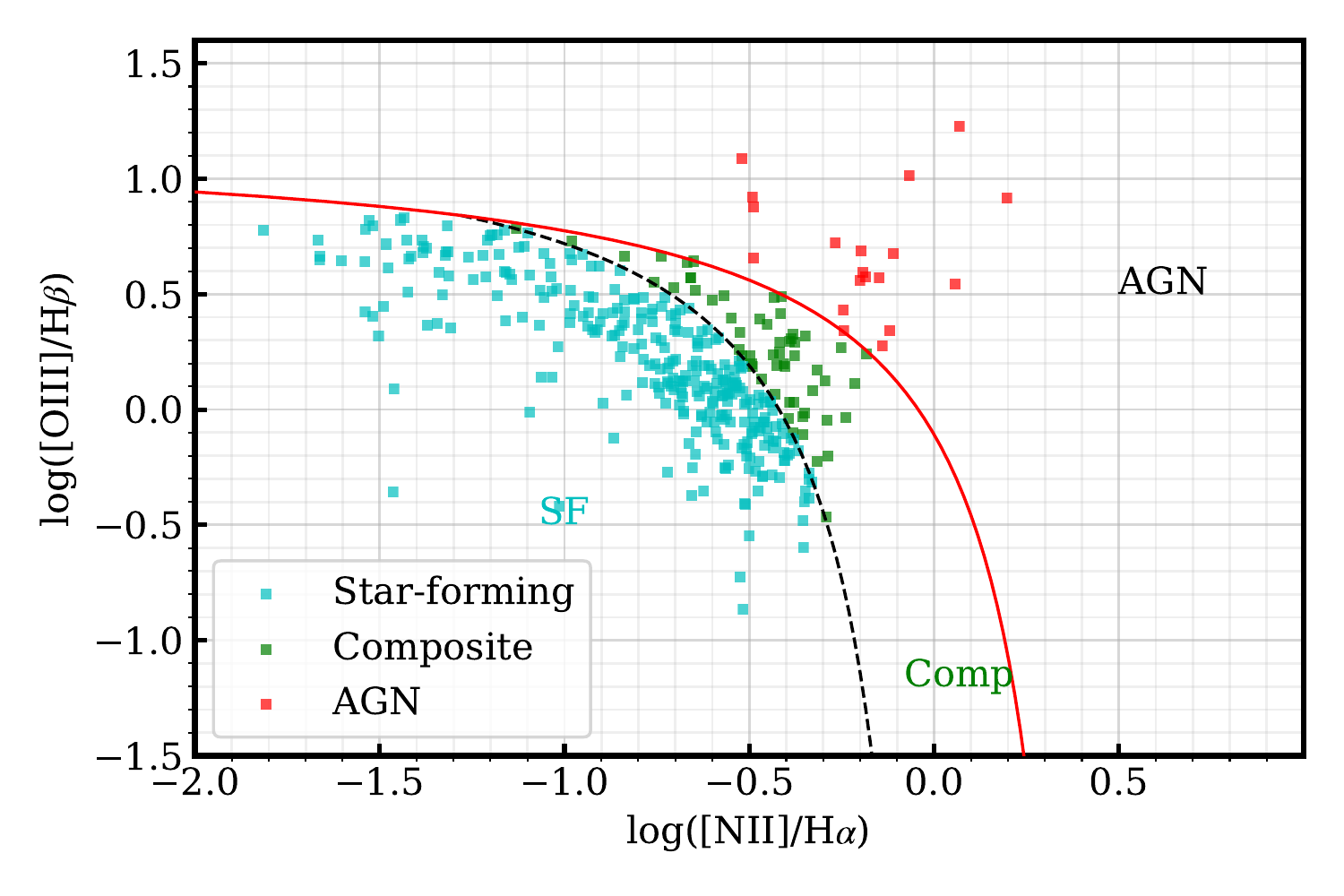}
      \caption{BPT diagram for the selected sources. Only 'SF' and `comp' sources are kept for further analysis. }
         \label{fig:bpt}
  \end{figure}

Using the massive amount of spectra available in the LAMOST database \citep{1996ApOpt..35.5155W,2004ChJAA...4....1S,2012RAA....12.1243L}, we cross-match the photometry sources with LAMOST DR10\footnote{\url{http://www.lamost.org/dr10/}} with a searching radius of 3'' and have obtained 2847 spectra that match these positions.
The goal of this work is to investigate the metallicity of the ionized gas in these low-$z$ galaxies, the AGN components will bias the predictions from the photoionization models.
Therefore, we remove the sources that show broad Balmer emission lines and display the [Mg~\textsc{ii}]$\lambda$2800 line if located within the wavelength coverage to eliminate AGNs from our sample.
Furthermore, based on the emission line ratios, we calculate the BPT diagram as in Figure \ref{fig:bpt} and keep the sources classified as `SF' or `comp' for further analysis in this work. 
We select the ones that have significant [O~\textsc{iii}]$\lambda$5007 emission lines with $\ge 2\sigma$ detection.
Thus, we have obtained 402 new spectra with 346 new sources in this work.  

\subsubsection{Flux calibration}
Following the procedures in \citet{2018MNRAS.474.1873W,2022ApJ...927...57L},
we re-calibrate the LAMOST spectra according to the SDSS $gri$ photometry.  
LAMOST spectra are convolved with the SDSS $gri$ filters \citep{1996AJ....111.1748F} in the observed frame to obtain the synthetic magnitudes for these three bands and we calculate the difference of the synthetic magnitudes with the observed SDSS photometric magnitudes.  
A zeroth-order or first-order polynomial is used to fit the magnitude difference array and apply this correction to the LAMOST spectra. 

 \subsection{Comparison sample}
To check whether there is redshift evolution of the relations discussed in Section \ref{sec:discussions}, there are also measurements at different redshift ranges included in this work for comparison.
At $z\sim 0$, we use the measurements from the composite spectra of SDSS emission line galaxies in \citet{2013ApJ...765..140A}.

At higher redshifts, we make use of two stacked measurements. 
\citet{2015ApJ...798...29Z} used stacked low-resolution near-IR grism spectra from the \textit{Hubble Space Telescope (HST)} and obtained composite measurements of $z\sim 2$ low-mass (median $\rm 10^9~M_\odot$) galaxies.
We include the measurements of five stellar mass bins in the composite spectra.
\citet{2020ApJ...902L..16J} measure the doubly ionized Neon ([Ne~\textsc{iii}]$\lambda3869$) for $z\sim2$ galaxies in the MOSFIRE DEEP Evolution Field (MOSDEF) survey \citep{2015ApJS..218...15K}.
We use the measurements from the stacked spectra in four stellar mass bins, which have no requirements of the [Ne~\textsc{iii}]$\lambda3869$ detection.

The revolutionary \textit{James Webb Space Telescope (JWST)} and its near-IR spectrograph NIRSpec \citep{2022A&A...661A..81F,2022A&A...661A..80J} has opened the new possibility of investigating the chemical abundances at higher-$z$.
Among the 35 objects detected in the field of the lensed galaxies of cluster SMACS J0723.3-7327 \citep{2001ApJ...553..668E,2007ApJ...661L..33E,2010MNRAS.407...83E,2012MNRAS.420.2120M} from the Early Release Observations (ERO) of \textit{JWST}, there are three sources at $z\sim 7.7-8.5$.
These three sources show similar rest-frame optical spectra as the strong [O~\textsc{iii}]$\lambda$5007 emission-line compact galaxies in this sample, which is worth to be compared with.
We use the measurements in \citet{2023MNRAS.518..425C} in this work.
  
\section{Physical Parameters}
\label{sec:para}
\subsection{Multi-wavelength spectral energy distribution fitting}
We cross-match the sources with the \textit{Galaxy Evolution Explorer (GALEX)} \citep{Bianchi_2017} to obtain the NUV photometry and cross-match the sources with the AllWISE Multiepoch Photometry Table  \citep{2010AJ....140.1868W,Wright_2019-os} to obtain the $W_1$ to $W_4$ photometry.
For the stellar mass measurement, we fit the sources with multiwavelength photometry with CIGALE \citep{2005MNRAS.360.1413B,2019A&A...622A.103B} combining the GALEX NUV, SDSS $ugriz$, and AllWISE $W_1$ to $W_4$ photometry.  
For the configuration of the fitting, we use the delayed$-\tau$ star formation history, BC03 stellar population models \citep{2003MNRAS.344.1000B}, Chabrier IMF \citep{2003PASP..115..763C}, nebular emission lines, the dust attenuated modified starburst model, the dust emission model from \citet{2012MNRAS.425.3094C}, and the \citet{2006MNRAS.366..767F} AGN model.

\subsection{Emission line flux, equivalent width, and velocity dispersion}
\begin{figure}[!h]
\centering
\subfloat{\includegraphics[width=0.45\hsize]{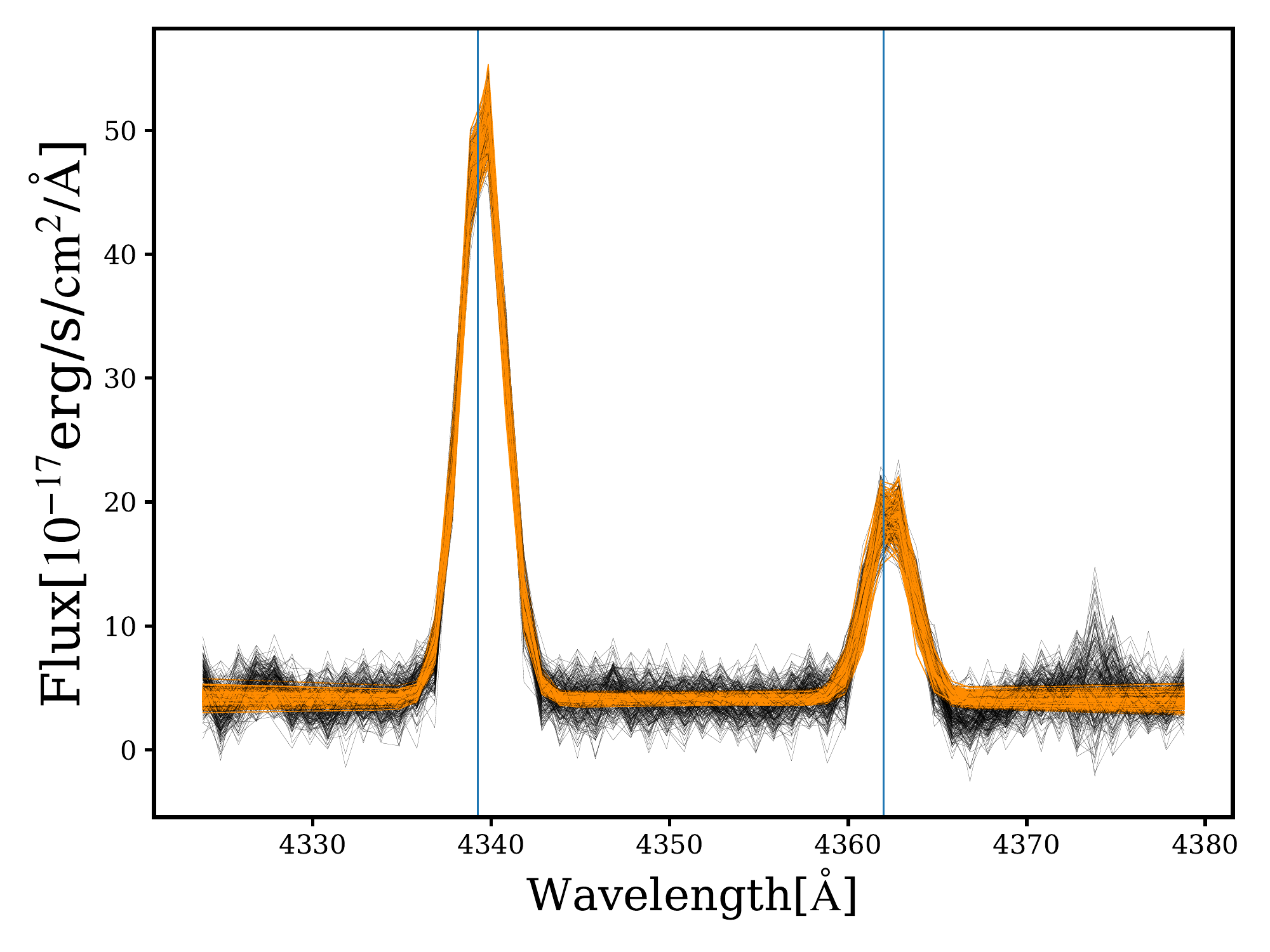}}
\subfloat{\includegraphics[width=0.45\hsize]{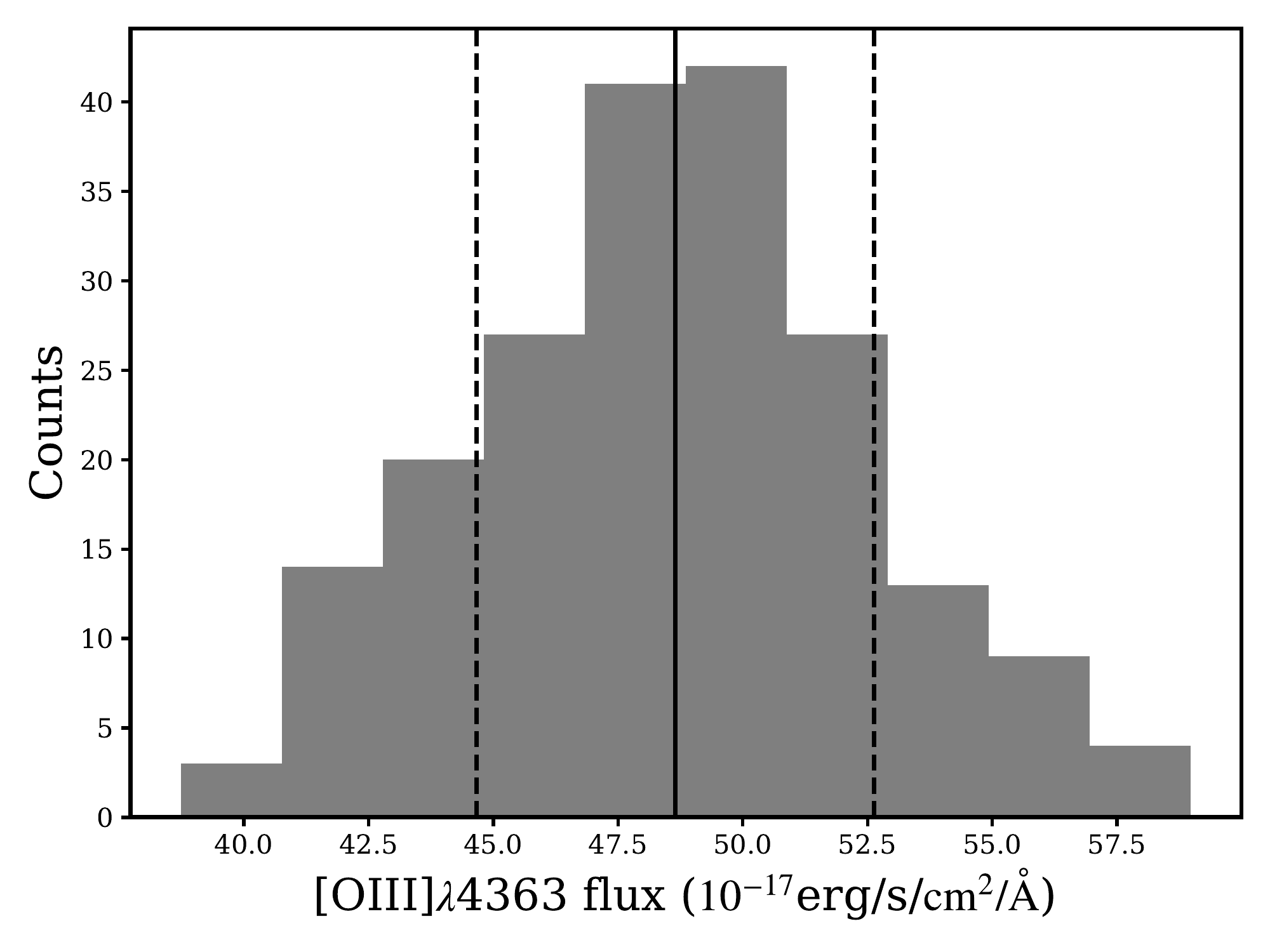}}
\caption{
The H$\gamma$ and [O {\sc iii}]$\lambda$4363 emission lines fitted with the Gaussian profiles (left panel) and histogram of the measurement of each realization (right panel).  
In the left panel, the flux of each realization is marked with a black line, and the fitted Gaussian profile is marked with an orange curve.
In the right panel, the mean value of the distribution is marked with the black solid line, and the 1$\sigma$ region is marked with the dashed lines.}
\label{fig:OIII4363_measurement}
\end{figure}

We fit the emission lines with \texttt{LMFIT} \citep{2016ascl.soft06014N}.
For emission line error estimation, we use the Monte Carlo method. 
For each instance, we perturb the flux at each wavelength sampling point 200 times by running a Gaussian distribution centered on the flux measurement with a width set by the flux error. 
The mean value of the resulting emission line flux is the resulting flux measurement, and the standard deviation marks the flux error.
An example demonstrating the measurement of [O~\textsc{iii}]$\lambda$4363 line is in Figure \ref{fig:OIII4363_measurement}.

\begin{figure}[!h]
  \centering
  \includegraphics[width=\hsize]{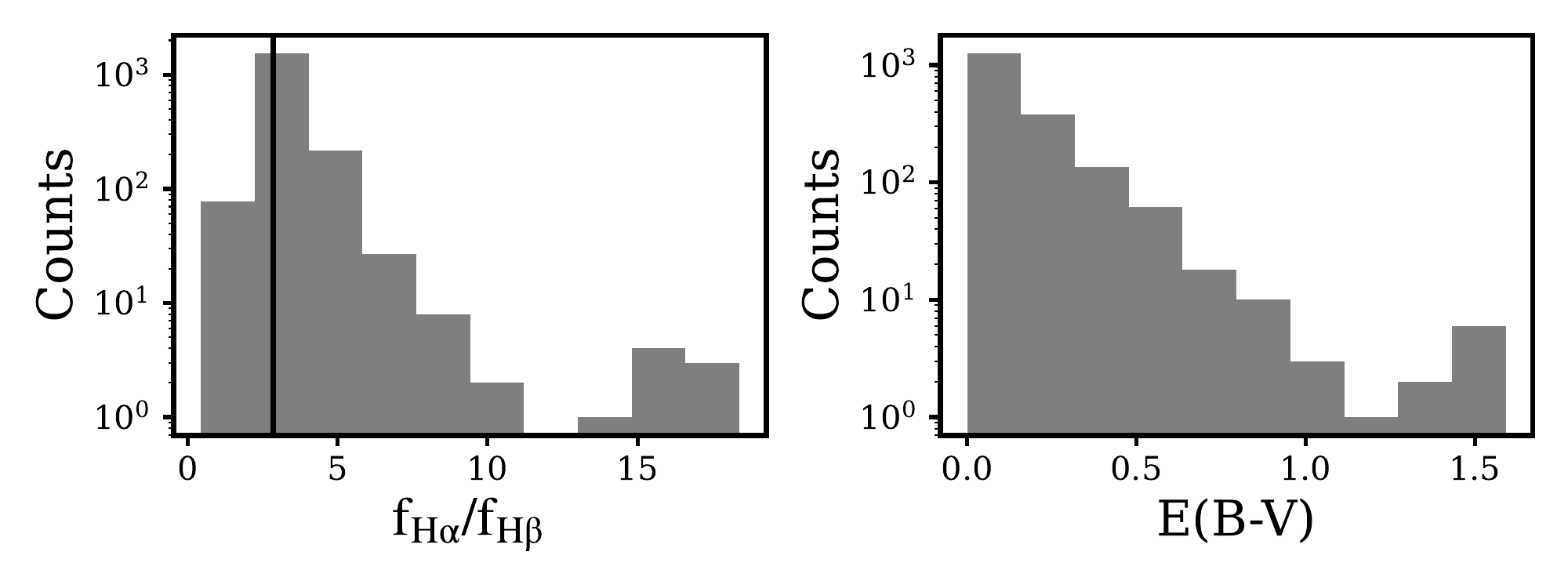}
      \caption{
      The distribution of the flux ratio of $\rm H\alpha/H\beta$ (left panel) and the E(B-V) (right panel).}
         \label{fig:color_excess}
  \end{figure}

We measure the color excess of the galaxies from the flux ratio of the H$\alpha$ to H$\beta$ assuming the case B recombination with the intrinsic line ratio of 2.86.
The color excess from the flux ratio is calculated with
\begin{eqnarray}
  \rm   E(B-V)_{gas} = \frac{\log_{10}[(f_{H\alpha}/f_{H\beta})/2.86]}{0.4\times [k(H\beta) - k(H\alpha)]},
\end{eqnarray}
where $ k(\rm H\alpha)=3.33$ and $k(\rm H\beta)=4.6$ as in \citet{2019ApJ...872..145J}.
If the flux ratio is less than 2.86, the color excess is considered to be set to 0.
The distribution of the flux ratios and E(B-V) are displayed in Figure \ref{fig:color_excess}. 
All of the emission line flux measurements are corrected with the extinction correction using the \citet{calzetti2000dust} extinction law, with the color excess measured from the Balmer decrement.

\begin{figure*}[ht]
  \centering
  \includegraphics[width=\hsize]{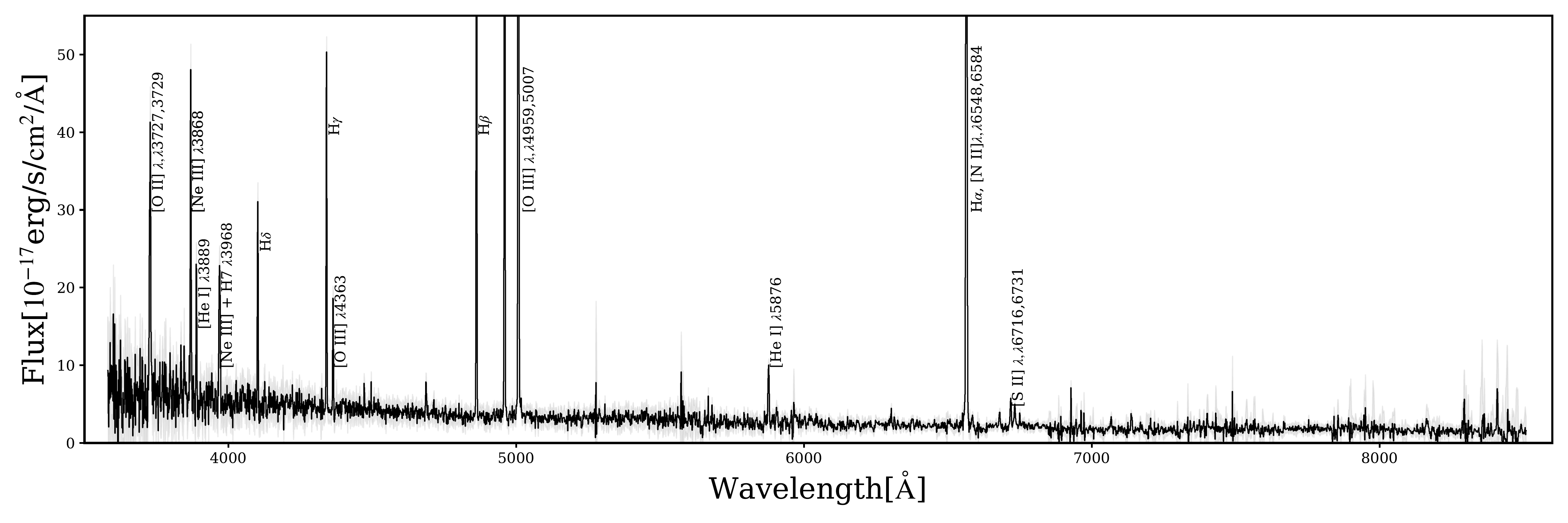}
      \caption{An example spectrum of the strong [O~\textsc{iii}]$\lambda$5007 emission-line compact galaxy. 
      The emission lines used for following metallicity and kinematics measurements and discussions are marked at different wavelengths.
     The grey-shaded regions mark the flux error.
      }
         \label{fig:277}
  \end{figure*}

We demonstrate an example of the spectrum with the emission lines marked in Figure \ref{fig:277}.

Specifically, we also calculate the velocity dispersion of the [O~\textsc{iii}]$\lambda5007$ emission line.
To calculate the intrinsic velocity dispersion of the sources, we subtract the instrumental and thermal broadening of the sources as in \citet{2019MNRAS.489.1787B,2020MNRAS.494.3541H}.
Assuming an electron temperature $\rm T_e=1.2\times 10^4 \rm K$ \citep{2010ApJ...715L.128A,2012ApJ...754L..22A}, the typical thermal broadening for these galaxies are $\rm \sigma_{\rm ther}=\frac{c\lambda_{\rm em}}{\lambda_{\rm obs}} \sqrt{\frac{k_BT_e}{m_{\rm ion}c^2}}$.
The thermal broadening of a galaxy at $z=0.26$ is about 2.50 km s$^{-1}$.  
We refer to the values given in \citet{2014RAA....14.1234S} for the instrumental FWHM as an average value of 3.5 \r{A} for LAMOST spectrographs.

\begin{figure}[ht]
  \centering
  \includegraphics[width=\hsize]{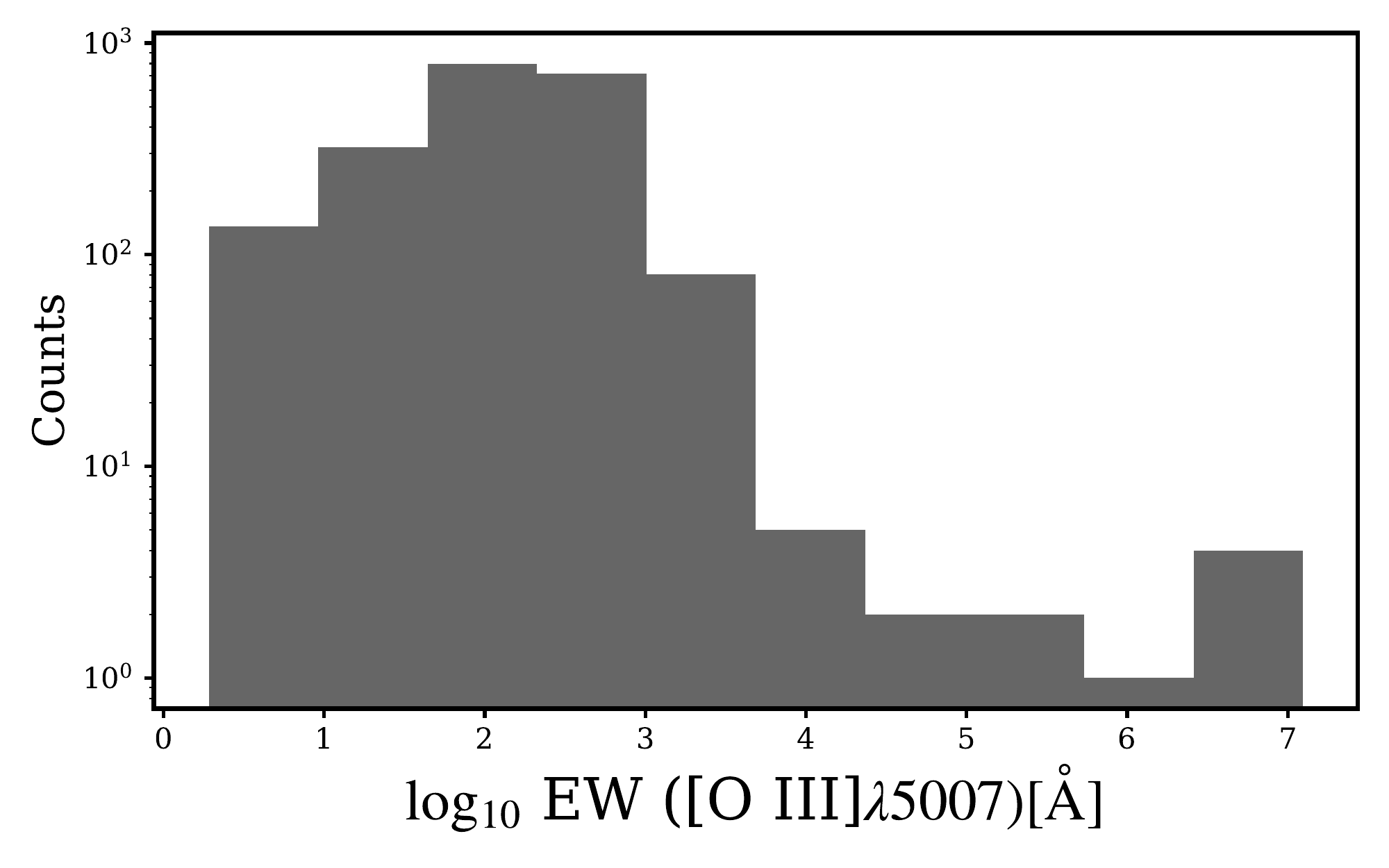}
      \caption{
      The distribution of the rest-frame EWs of the [O~\textsc{iii}]$\lambda5007$ emission lines.
      }
         \label{fig:EW}
  \end{figure}

The rest-frame equivalent widths (EWs) of the [O~\textsc{iii}]$\lambda5007$ emission lines are measured to be compared with other work. 
Figure \ref{fig:EW} demonstrates the distribution for the measured EWs.\footnote{Due to the faint continuum, some of the EWs are extremely large, thus the measured result for EW$>10^5$\r{A} is not completely reliable.}

All the above measurements are compiled into a catalog \url{https://nadc.china-vo.org/res/r101238}, and the descriptions of the catalog are listed in Table \ref{tab:table_description}. 

\begin{deluxetable*}{@{\extracolsep{6pt}}ll@{}}[!ht]
\tablecaption{Catalog: Table Description\label{tab:table_description}}
\tablehead{  \colhead{\textbf{Name}}  & \colhead{\textbf{Description}}}
\startdata
combined\_obsid & unique object id in LAMOST database\\
combined\_lmjd  &  Local Modified Julian Day (LMJD )of the spectroscopic observation\\
combined\_planid & spectroscopic plan identifications\\
combined\_spid &  spectrograph identification \\
combined\_fiberid & fiber number of the spectrum\\
combined\_objtype & object type from input catalog\\
combined\_class & class of object\\
combined\_z & spectroscopic redshift\\
combined\_tfrom & Target catalog\\
combined\_ra &  R.A in decimal degrees (J2000)\\
combined\_dec & declination in decimal degrees (J2000)\\
\hline 
objID & unique SDSS identifier \\
type & type classification of the object (0=UNKNOWN, 3=galaxy, 6=star)\\
clean &  clean photometry flag (1=clean, 0=unclean).\\
cModelMag\_? & DeV+Exp magnitude, ``?" represents each of the $ugriz$ SDSS bands\\
cModelMagErr\_? & DeV+Exp magnitude error, ``?" represents each of the $ugriz$ SDSS bands\\
petroRad\_r & Petrosian radius in the $r-$band\\
petroR50\_r & Radius containing 50\% of Petrosian flux in the $r-$band\\
petroR90\_r & Radius containing 90\% of Petrosian flux in the $r-$band\\
\hline
OII\_flux(\_err) & [O~\textsc{ii}]$\lambda\lambda$3727,3729 line flux and flux error in units of $\rm 10^{-17} erg ~s^{-1} cm^{-2}$\\
NeIII3869\_flux(\_err) & [Ne~\textsc{iii}]$\lambda$3869 line flux and flux error in units of $\rm 10^{-17} erg ~s^{-1} cm^{-2}$\\
OIII4363\_flux(\_err) & [O~\textsc{iii}]$\lambda$4363 line flux and flux error in units of $\rm 10^{-17} erg ~s^{-1} cm^{-2}$\\
Hb\_flux(\_err) & H$\beta$ line flux and flux error in units of $\rm 10^{-17} erg ~s^{-1} cm^{-2}$\\
OIII4959\_flux(\_err) & [O~\textsc{iii}]$\lambda$4959 line flux and flux error in units of $\rm 10^{-17} erg ~s^{-1} cm^{-2}$\\
OIII5007\_flux(\_err) & [O~\textsc{iii}]$\lambda$5007 flux and flux error in units of $\rm 10^{-17} erg ~s^{-1} cm^{-2}$\\
NII6548\_flux(\_err) & [N~\textsc{ii}]$\lambda$6548 flux and flux error in units of $\rm 10^{-17} erg ~s^{-1} cm^{-2}$\\
Ha\_flux(\_err) & H$\alpha$ line flux and flux error in units of $\rm 10^{-17} erg ~s^{-1} cm^{-2}$\\
NII6584\_flux(\_err) & [N~\textsc{ii}]$\lambda$6584 flux and flux error in units of $\rm 10^{-17} erg ~s^{-1} cm^{-2}$\\
SII6716\_flux(\_err) & [S~\textsc{ii}]$\lambda$6716 flux and flux error in units of $\rm 10^{-17} erg ~s^{-1} cm^{-2}$\\
SII6731\_flux(\_err) & [S~\textsc{ii}]$\lambda$6731 flux and flux error in units of $\rm 10^{-17} erg ~s^{-1} cm^{-2}$\\
\hline
EW\_OIII5007 & Equivalent width of the [O~\textsc{iii}]$\lambda$5007 emission line in units of \r{A}  \\
ne & electron number density in units of $\rm cm^{-3}$\\
cigale\_mass & stellar mass in units of $\rm M_{\odot}$ \\
sigma\_5007 & velocity dispersion of the [O~\textsc{iii}]$\lambda5007$ emission line, in units of \r{A} \\
\hline
\enddata
\end{deluxetable*}

\section{Results}
\label{sec:results}

\subsection{Direct-$T_e$ method gas-phase metallicity}
   
With the auroral [O~\textsc{iii}]$\lambda4363$ line, the gas-phase metallicity could be estimated in a direct-$T_e$ method \citep{1984ASSL..112.....A,2006A&A...448..955I}.
This approach assumes the electron temperatures for $\rm O^+$ and $\rm O^{++}$ in a two-zone photoionization model: the high-ionization zone traced by the $\rm O^{++}$ and the low-ionization one traced by the $\rm O^{+}$.
We calculate the $\rm O^{++}$ electron temperature with the following equation (as in \citet{2006A&A...448..955I} Eqs (1) and (2)):
\begin{eqnarray}
 \rm   t = \frac{1.432}{\log[(\lambda4959 + \lambda 5007)/\lambda4363]-\log C_T},
\end{eqnarray}
where $\rm t=10^{-4}T_e([\rm O~\textsc{iii}])$, and 
\begin{eqnarray}
\label{eq:CT}
\rm C_T = (8.44-1.09t+0.5t^2-0.08t^3)\frac{1+0.0004x}{1+0.044x},
\end{eqnarray}
where $\rm x=10^{-4}n_et^{-0.5}$.
As addressed in \citet{2013ApJ...765..140A,2017MNRAS.465.1384C}, at low-metallicity environments, the $\rm O^{++}$ abundance is dominant over that of $\rm O^{+}$.
Besides, there is no direct temperature diagnostic for the low-ionization zone, the electron temperature of [O~\textsc{ii}] for the low-metallicity situation is estimated with \citet{2006A&A...448..955I} as:
\begin{eqnarray}
  \rm  T_e([{\rm O~\textsc{ii}}]) &=&\rm  -0.577 + T_e([{\rm O~\textsc{iii}}])\times \nonumber\\
   & &\rm (2.065-0.498T_e([{\rm O~\textsc{iii}}])).
\end{eqnarray}
This is also the method that \citet{2019ApJ...872..145J} use for calculation, and they state that their measurement of the oxygen abundance depends little on the relation between $\rm t_2$ and $\rm t_3$.
Similarly, we use the term $\rm t_2 = 10^{-4}T_e([{\rm O~\textsc{ii}}])$ and $\rm t_3 = 10^{-4}T_e([{\rm O~\textsc{iii}}])$ for clarity.

\begin{figure}[!ht]
  \centering
  \includegraphics[width=\hsize]{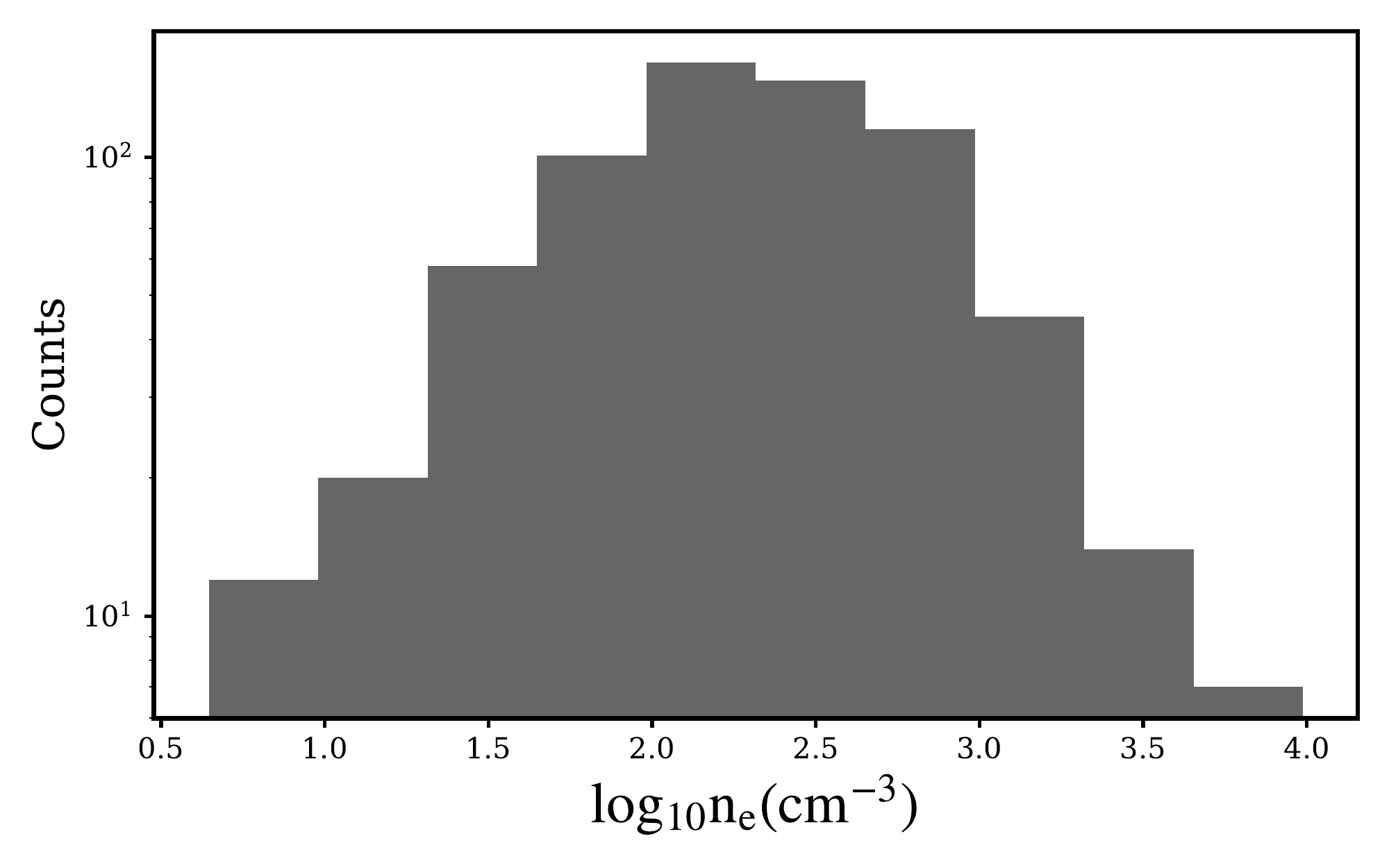}
      \caption{The distribution of the derived $\rm n_e$.}
         \label{fig:hist_ne}
  \end{figure}

We follow the treatment as in \citet{2019ApJ...872..145J} using the [S \textsc{ii}]$\lambda,\lambda$6716,6731 lines to estimate the electron density $\rm n_e$.  
For 1052 sources with both $2\sigma$ detections for the [S \textsc{ii}]$\lambda$6716 and [S \textsc{ii}]$\lambda$6731 emission lines, the flux ratio is defined as $\rm R(n_e) = \frac{\lambda 6716}{\lambda 6731}$, and the electron density is estimated\footnote{For $\rm R(n_e) > 1.43$, quoting the low-density limits below the flux ratio is unrealistic, thus it is merged with the situation with $\rm 0.51 \le R(n_e) \le 1.43$.}:
\begin{equation}
\label{eq:ne}
\rm n_e (R)=
\begin{cases}
      \rm \frac{ab - Rc}{R-a}, &  \rm R \ge 0.51  \\
      100, & \rm  otherwise\\
    \end{cases}
\end{equation}
in units of $\rm cm^{-3}$, where $a=0.4441,b=2514,c=779.3$.
The derived distribution of $\rm n_e$ is in Figure \ref{fig:hist_ne}.
As addressed in \citet{2019ApJ...872..145J}, under $\rm n_e=10, 100 $ or $\rm 10^3~cm^{-3}$, the results do not vary much.

With Eqs. (3) and (5) in \citet{2006A&A...448..955I}, we calculate the ionic abundances as follows:
\begin{eqnarray}
    \rm  12+\log \frac{O^+}{H^+} &=& \rm \log \frac{\lambda 3727}{H\beta} + 5.961 + \frac{1.676}{t_2} - 0.40 \log t_2 \nonumber\\
   & &\rm - 0.034 t_2 + \log(1+1.35x),
\end{eqnarray}
and 
\begin{eqnarray}
  \rm  12+\log \frac{O^{++}}{H^+} &=& \rm \log \frac{\lambda 4959 + \lambda5007}{H\beta} + 6.200 + \frac{1.251}{t_3} \nonumber\\
    & &\rm - 0.55\log t_3 - 0.014 t_3,
\end{eqnarray}
Since the majority of the ions of oxygen are $\rm O^{+}$ and $\rm O^{++}$, we use $\rm \frac{O}{H} = \frac{O^+}{H} + \frac{O^{++}}{H}$ to determine the oxygen abundance, neglecting the contribution from higher ionization states.

Among the total sample of the strong [O~\textsc{iii}]$\lambda$5007 emission-line compact galaxies, there are 450 sources with 2$\sigma$ [O~\textsc{iii}]$\lambda4363$ detection. 
Only the sources where the $\rm t_3$ range from 5 000 to 20 000 K, as noted in \citet{2006A&A...448..955I}, are selected for metallicity discussion, which counts to 372.
For this sub-sample, we estimate the gas-phase metallicity and use them for the following metallicity measurements and discussions.

\subsection{Strong-line calibration}

\begin{figure*}[!ht]
  \centering
  \includegraphics[width=\hsize]{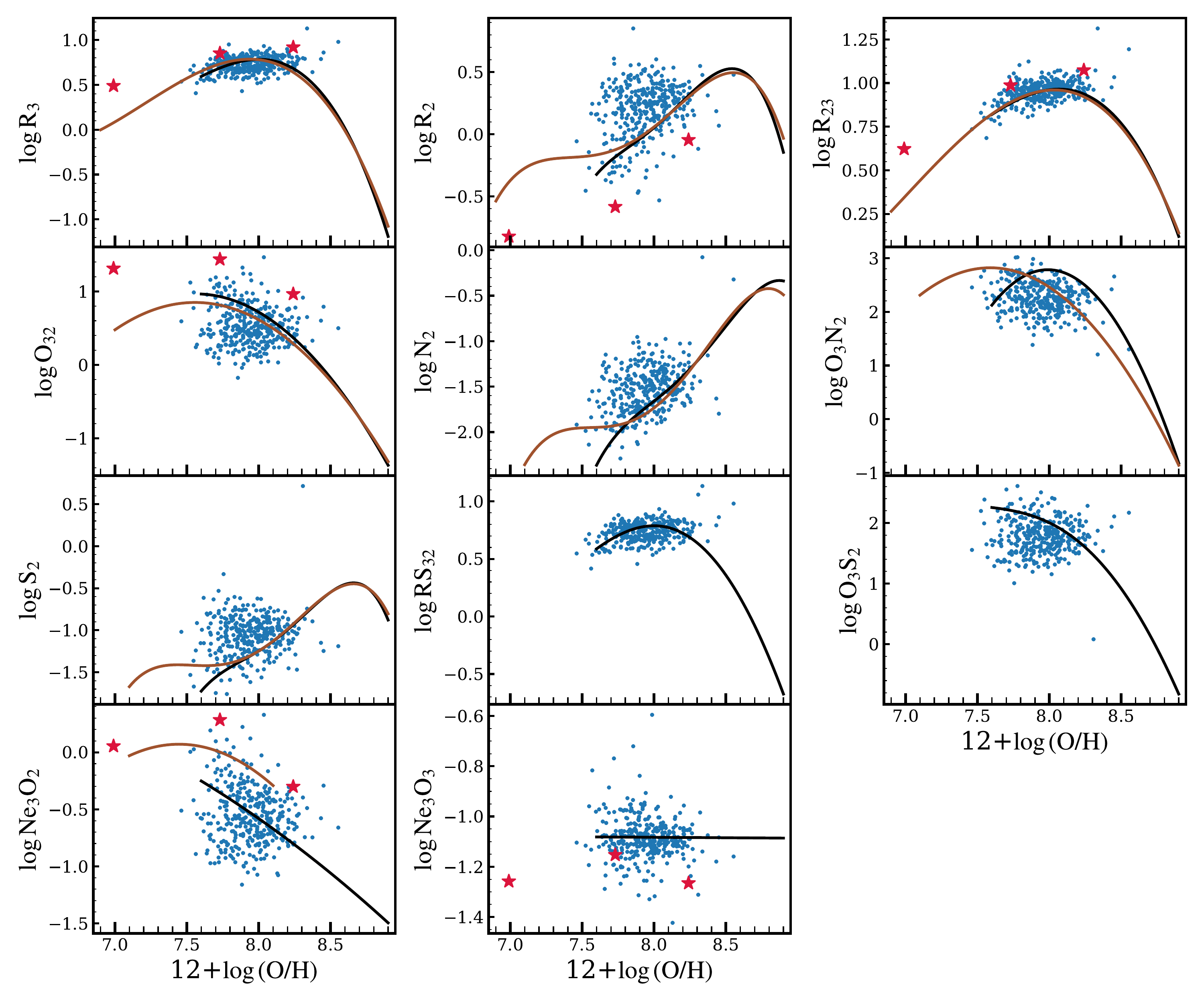}
      \caption{The set of strong-line calibrations from \citet{2008A&A...488..463M,2015ApJ...813..126J,2020MNRAS.491..944C} (black curve) and \citet{2022ApJS..262....3N} (sienna curve).
      The x-axis marks the metallicity obtained from the direct-$T_e$ method.
      The distribution of the strong [O~\textsc{iii}]$\lambda$5007 emission-line compact galaxies is displayed with blue scatter points.
      The crimson stars mark the three galaxies from \textit{JWST} at $z\sim 7.7-8.5$ with the measurements from \citet{2023MNRAS.518..425C}.
      }
         \label{fig:strong-line_diag}
  \end{figure*}

We use different strong-line calibrations and verify whether these strong [O~\textsc{iii}]$\lambda$5007 emission-line compact galaxies follow the derived scaling relations of the strong-line calibrations.  
The definitions of the strong-line calibration are listed in Table \ref{tab:strong-line}.
Comparing with the strong-line calibrated polynomial forms from \citet{2008A&A...488..463M,2015ApJ...813..126J,2020MNRAS.491..944C}\footnote{The Ne3O2 calibration is from \citet{2008A&A...488..463M}, and the Ne3O3 calibration is from \citet{2015ApJ...813..126J}.} as in Figure \ref{fig:strong-line_diag}, we find that the strong [O~\textsc{iii}]$\lambda$5007 emission-line compact galaxies agree with these strong-line diagnostics.
This set of calibration is valid in $\rm 7.6 \le 12 + \log(O/H) \le 8.9$.

For the low-metallicity regime, we also include the strong-line calibration from \citet{2022ApJS..262....3N}.\footnote{We adopt Ne3O2 relation for large EW samples in \citet{2022ApJS..262....3N}.} 
The optical-line gas metallicity diagnostics are established by the combination of local SDSS galaxies \citep{2017MNRAS.465.1384C} and the largest compilation of extremely metal-poor galaxies (EMPGs) identified by the Subaru EMPRESS survey \citep{2020ApJ...898..142K}.
This set of calibrations reaches the lower metallicity roughly down to $\rm 12+\log(O/H)\sim 6.9$.

\begin{deluxetable}{@{\extracolsep{6pt}}ll@{}}[!ht]
\tablecaption{Strong-line calibration\label{tab:strong-line}}
\tablehead{  
\colhead{\textbf{Notation}}  & \colhead{\textbf{Line Ratio}}}
\startdata
$\rm R_2$ & [O {\sc ii}]$\lambda\lambda 3727,3729$/H$\beta$\\
$\rm R_3$ & [O {\sc iii}]$\lambda5007$/H$\beta$\\
$\rm R_{23}$ & ([O {\sc ii}]$\lambda,\lambda 3727,3729$ + [O {\sc iii}]$\lambda\lambda 4959,5007$)/H$\alpha$\\
$\rm O_{32}$ & [O {\sc iii}]$\lambda5007$/[O {\sc ii}]$\lambda\lambda 3727,3729$\\
$\rm N_2$ & [N {\sc ii}]$\lambda6584$/H$\alpha$\\
$\rm O_3N_2$ & ([O {\sc iii}]$\lambda5007$/H$\beta$)/([N {\sc ii}]$\lambda6584$/H$\alpha$)\\
$\rm S_2$ & [S {\sc ii}]$\lambda\lambda 6717,6731$/H$\alpha$\\
$\rm RS_{23}$ & [O {\sc iii}]$\lambda5007$/H$\beta$ + [S {\sc ii}]$\lambda\lambda 6717,6731$/H$\alpha$\\
$\rm O_3S_2$ & ([O {\sc iii}]$\lambda5007$/H$\beta$)/([S {\sc ii}]$\lambda\lambda 6717,6731$/H$\alpha$)\\
$\rm Ne_3O_2$ & [Ne {\sc iii}]$\lambda3869$/[O {\sc ii}]$\lambda\lambda 3727,3729$\\
$\rm Ne_3O_3$ & [Ne {\sc iii}]$\lambda3869$/[O {\sc iii}]$\lambda5007$\\
\hline
\enddata
\end{deluxetable}

\section{Discussions}
\label{sec:discussions}
\subsection{Mass-metallicity relation}
\begin{figure}[!h]
  \centering
  \includegraphics[width=\hsize]{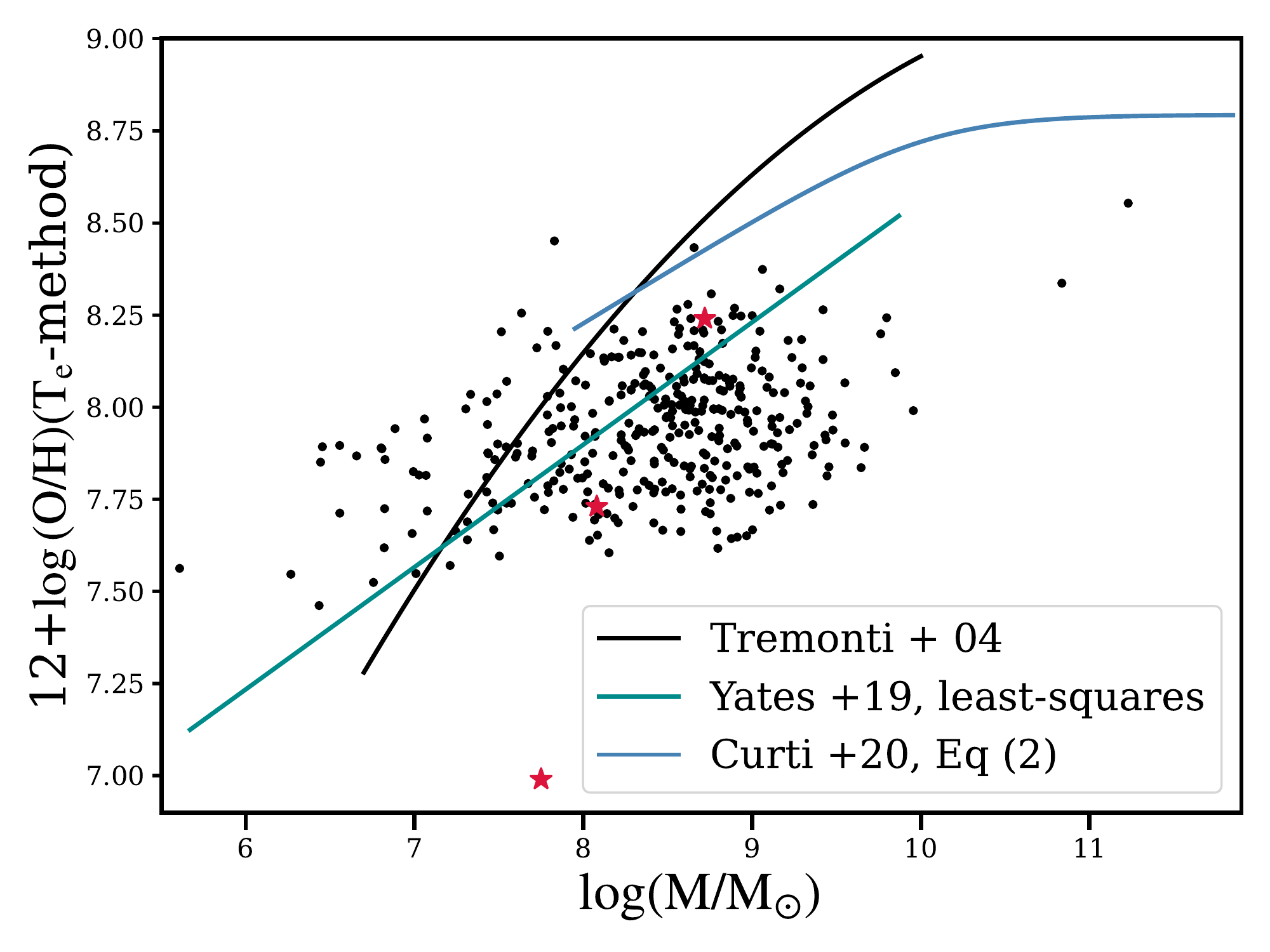}
      \caption{The mass metallicity relation of the strong [O~\textsc{iii}]$\lambda$5007 emission-line compact galaxies.
      Different mass-metallicity relations from \citet{2004ApJ...613..898T,2020A&A...634A.107Y,2020MNRAS.491..944C} are marked with curves with the corresponding colors as noted in the legend.
      The crimson stars mark the three galaxies from \textit{JWST} at $z\sim 7.7-8.5$ with the measurements from \citet{2023MNRAS.518..425C}.
      }
         \label{fig:mass-metal}
  \end{figure}

With the gas-phase metallicity determined from the direct-$T_e$ method, we demonstrate in Figure \ref{fig:mass-metal} that the metallicity increases with the stellar mass and falls below several $T_e$-calibrated MZRs.

\citet{2004ApJ...613..898T} and later works \citep{2010MNRAS.408.2115M} adopt the calibration from different grids of photoionization models.
The MZR from \citet{2020A&A...634A.107Y} is based on the samples with auroral detections from the revised version of the classical $T_e$-based method.
Furthermore, \citet{2020MNRAS.491..944C} provide the MZR by fully anchoring the metallicity determinations for SDSS galaxies on the $T_e$ abundance employing the strong-line metallicity calibrations, which better captures the turnover mass.
The mass range is from $7.95<\log (\rm M/M_{\odot})  < 11.85$, which can be extrapolated to the low-mass regime.
The MZR of \citet{2020A&A...634A.107Y} is characterized by a lower normalization, which might be caused by the requirement of the [O~\textsc{iii}]$\lambda4363$ detection in the mass range of $5.67<\log (\rm M/M_{\odot})  < 9.87$.
As shown in Fig.4 in \citet{2020MNRAS.491..944C}, the galaxies in SDSS-DR7 with auroral detections locate below the MZR at low stellar masses. 
The majority of these strong [O~\textsc{iii}]$\lambda4363$ emission-line compact galaxies (the strong [O~\textsc{iii}]$\lambda$5007 emission-line compact galaxies) in this work are below the direct-$T_e$ method calibrated \citet{2004ApJ...613..898T,2020MNRAS.491..944C} mass-metallicity relation, and overlap the \citet{2020A&A...634A.107Y} relation.
This is reasonable because at fixed stellar mass, requiring the [O~\textsc{iii}]$\lambda4363$ detection preferentially selects the metal-poor galaxies.

\subsection{N/O vs O/H relation}

The total N/O ratio is not easy to measure since the [N~\textsc{iii}] lines are not readily observable, therefore, $\rm N^+/O^+$ are often used as a proxy of N/O abundance.
We determine the ratios following \citet{1992MNRAS.255..325P,2005A&A...437..849S} as :
\begin{eqnarray}
  \rm  \log \frac{N}{O} &=& \rm \log \frac{N^+}{O^+} \nonumber \\
  &=& \rm \log \frac{I([N~\textsc{ii}]\lambda\lambda 6548,6584)}{I([O~\textsc{ii}]\lambda\lambda3727,3729)} + 0.31  - \frac{0.726}{T_e([N~\textsc{ii}])} \nonumber\\
    & &\rm  - 0.02\log T_e([N~\textsc{ii}]) - \log \frac{1+1.35x}{1+0.12x},
\end{eqnarray}
where the $\rm T_e([N~\textsc{ii}]) = T_e([O~\textsc{ii}])$, and the definition of $\rm x$ is the same as in Equation \ref{eq:CT}.

\begin{figure*}[!ht]
  \centering
  \includegraphics[width=\hsize]{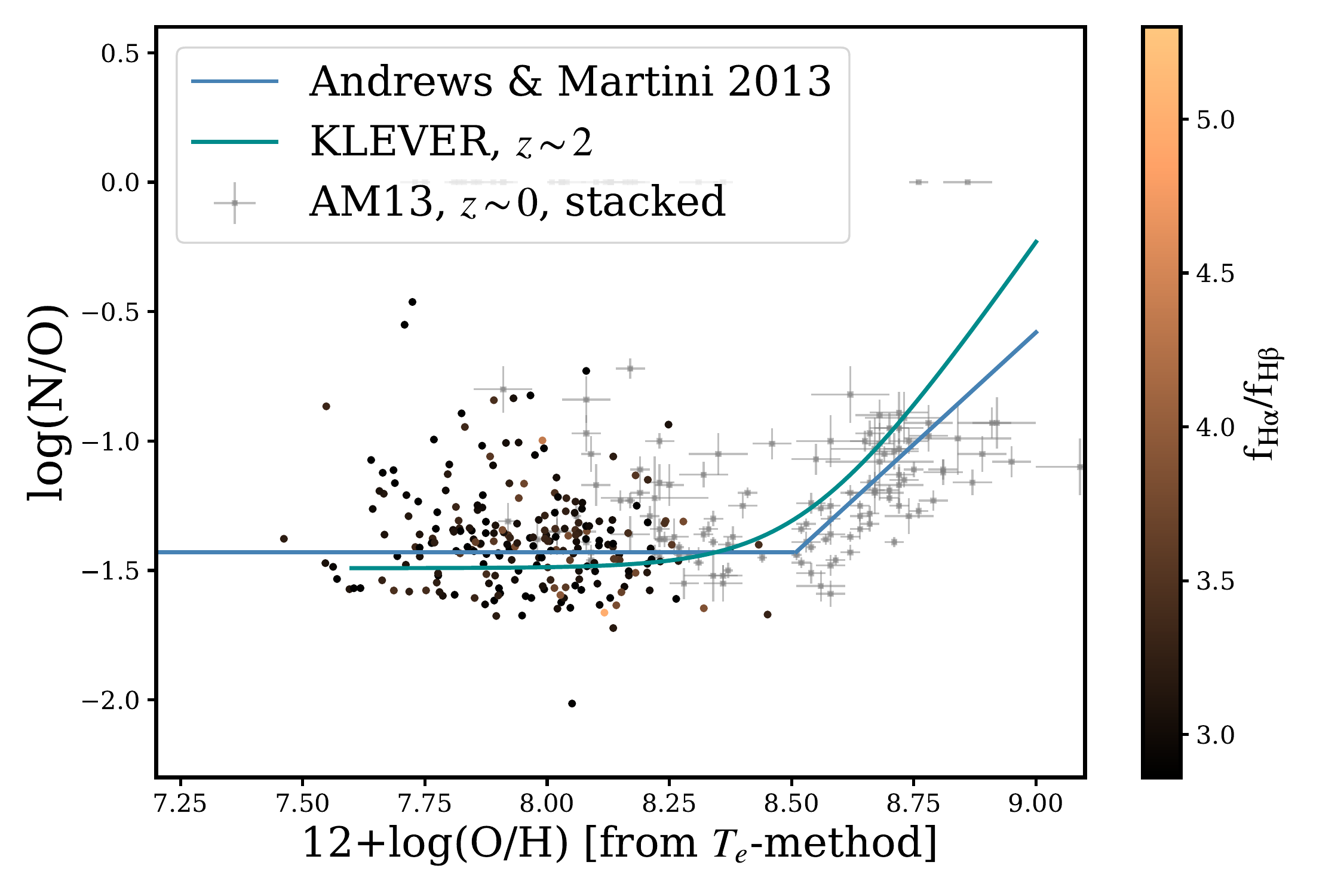}
      \caption{The N/O vs O/H relation of the strong [O~\textsc{iii}]$\lambda$5007 emission-line compact galaxies.
The colors of the scatter points mark the flux ratio of $\rm H\alpha$ to $\rm H\beta$.
    The measurements of the stacked spectra in bins of stellar mass and SFR from \citet{2013ApJ...765..140A} at $z\sim 0$ are marked with gray stars.
      The blue curve is the predictions from \citet{2013ApJ...765..140A}.
      The green curve is the fitting result from the KLEVER survey at $z\sim2$ \citep{2022MNRAS.512.2867H}.
      }
         \label{fig:NO-OH}
  \end{figure*}

The relation of the N/O vs O/H is displayed in Figure \ref{fig:NO-OH}.
Most of the samples are located in the horizontal range of the distribution with an approximately constant value, where the majority of the nitrogen is produced from the primary nucleosynthesis process. 
The colors of the scatter mark the flux ratio of $\rm H\alpha$ to $\rm H\beta$. 
It is well-known that there is a plateau in the N/O ratio at low metallicity \citep{1997nceg.book.....P,2016MNRAS.458.3466V,2022MNRAS.512.2867H}.
At $\rm 12+\log O/H = 8.5$ there is a transition that N/O increases with the metallicity, which is strongly influenced by the star formation efficiency \citep{2006MNRAS.372.1069M}.
With the evolution of the galaxies, low- to intermediate-mass stars produce secondary and primary nitrogen that increases the N/O ratio with the increase of metallicity at the same time. 
These strong [O~\textsc{iii}]$\lambda$5007 emission-line compact galaxies are mainly young galaxies with low metallicity.
The N/O values of the strong [O~\textsc{iii}]$\lambda$5007 emission-line compact galaxies almost have a constant value, indicated by the horizontal lines \citep{2013ApJ...765..140A,2022MNRAS.512.2867H} in Figure \ref{fig:NO-OH}.

As demonstrated in Figure \ref{fig:NO-OH}, a considerable number of samples show N/O value above the plateau at low metallicity.
However, the reason for this offset is not clear. 
No systematic trend of the flux ratio of $\rm H\alpha$ to $\rm H\beta$, as shown with the color of the scatter points, with this offset is observed. 
\citet{2010ApJ...715L.128A} claim similar trends of the Green Pea samples compared with the SDSS star-forming galaxies. 
The inflow of gas might explain this offset as well as the high SFR, compactness, and disturbed morphology.
\citet{2010ApJ...715L.128A} also add another explanation that this process is coupled with the selective metal-rich gas loss, driven by supernova winds.
However, based on IFU observations, \citet{2021ApJ...920L..46K} argue that the broad wings from local Green Pea analogs do not originate from stellar winds or supernovae, but are more likely driven by radiation.
It would be worth further investigation regarding the origin of this N/O offset at the low metallicity regime.

\citet{2016MNRAS.458.3466V} have modeled that the impact of changing IMF mainly is to shift the chemical evolution tracks along the metallicity axis, and the data from SDSS is best-fit with the Kroupa IMF \citep{2001MNRAS.322..231K}.
As demonstrated in Fig.8 in \citet{2016MNRAS.458.3466V}, if assuming the model with the Geneva stellar yields by taking the mass loss and rotation into consideration, the chemical evolution model predicted a dip in the N/O ratios.
Based on the observations of the metal-poor stars in \citet{2005A&A...430..655S}, Fig.2 in \citet{2005A&A...437..429C} demonstrates the N/O vs O/H plane, where there is also a significant scatter at low metallicity.
In Figure \ref{fig:NO-OH}, the scatter in N/O of the strong [O~\textsc{iii}]$\lambda$5007 emission-line compact galaxies in this work is comparable to that of the \citet{2013ApJ...765..140A} samples.

\subsection{Ne3O2 vs O32 relation and Ne3O2 vs mass relation}
Ne3O2 and O32 are both metallicity indicators \citep{2007MNRAS.381..125P,2007A&A...475..409S} as well as ionization parameter indicators \citep{2014ApJ...780..100L}.
The tight correlation between these two ratios is verified by different star-forming galaxies, which indicate similar evolution.
This is also checked for different metallicity as demonstrated in the Ne3O3 vs metallicity relation in Figure \ref{fig:strong-line_diag}, where Ne3O3 is almost flat for different metallicity ranges.
The strong [O~\textsc{iii}]$\lambda$5007 emission-line compact galaxies provide rich sources to verify the relations of the two ratios in the local Universe, especially at the high ionization regime.

\begin{figure*}[!ht]
  \centering
  \includegraphics[width=\hsize]{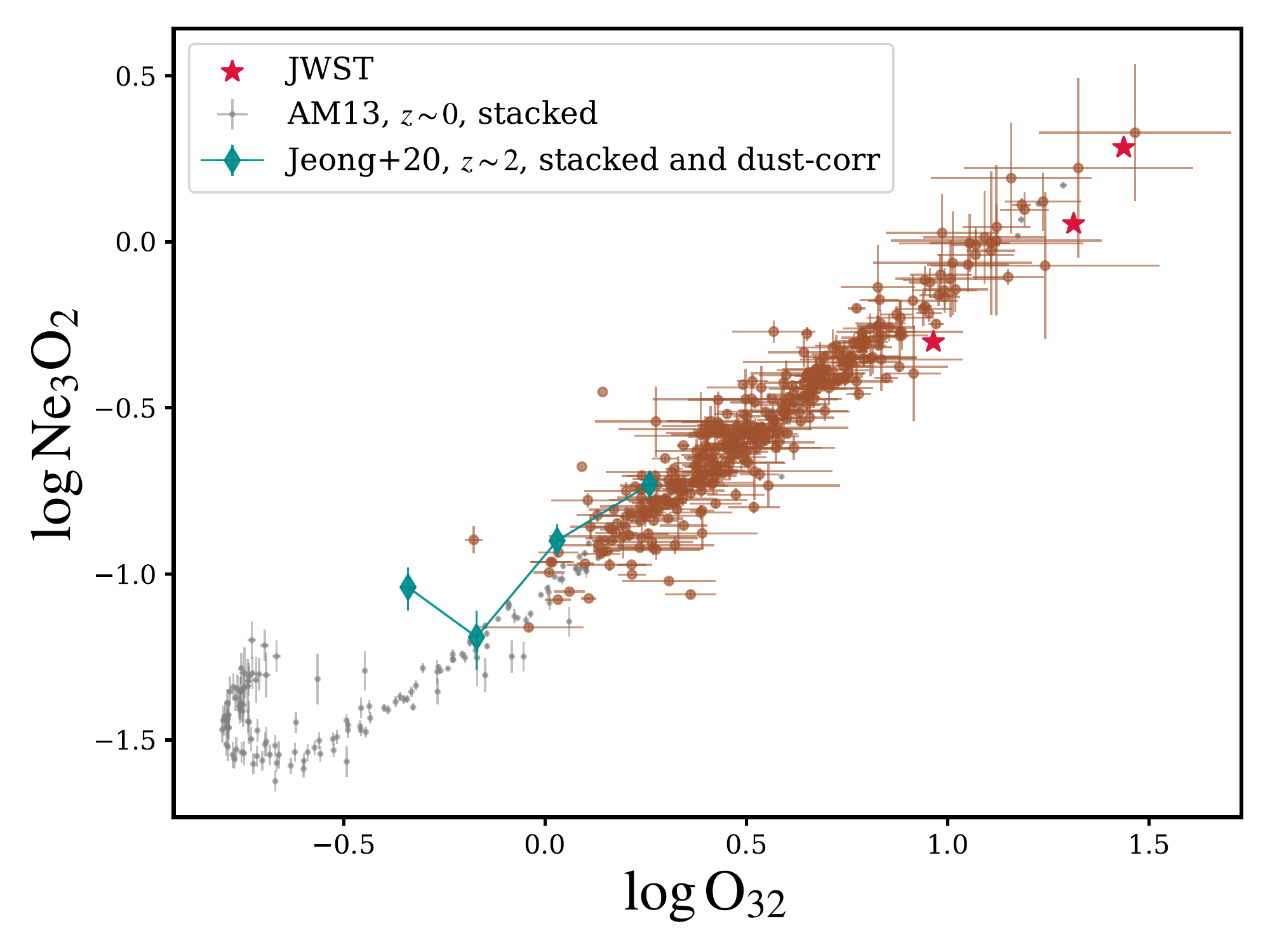}
      \caption{
      Ne3O2 vs O32 relation. 
      The \sout{pea samples} \textbf{strong [O~\textsc{iii}]$\lambda$5007 emission-line compact galaxies} are demonstrated with sienna error bars.
      The samples at $z\sim 0$ from \citet{2013ApJ...765..140A} are marked with gray stars.
      The stacked samples with the dust corrected from \citet{2020ApJ...902L..16J} are marked with dark cyan diamonds.
      The crimson stars mark the three galaxies from \textit{JWST}.
      }
         \label{fig:Ne3O2-O32}
  \end{figure*}

In Figure \ref{fig:Ne3O2-O32}, we demonstrate the relation of these two line ratios of the strong [O~\textsc{iii}]$\lambda$5007 emission-line compact galaxies in this work, the SDSS local star-forming galaxies from \citet{2013ApJ...765..140A}, the measurements of the stacked spectra in \citet{2020ApJ...902L..16J}, and three $z\sim 7.7-8.5$ sources from \textit{JWST}.
The two ratios follow tight linear relations, although the scatter is significant at the low ratio regime as predicted by the photoionization models \citep{2014ApJ...780..100L}. 
There is no significant redshift evolution of the linear relationship between these two parameters as seen from the measurements in \citet{2020ApJ...902L..16J,2023MNRAS.518..425C}.
At the low ionization regime, there are offsets between the \citet{2020ApJ...902L..16J} measurement and the local measurements from \citet{2013ApJ...765..140A}.
Due to the compactness of the strong [O~\textsc{iii}]$\lambda$5007 emission-line compact galaxies, they are less prone to the effects of the diffuse ionized gas (DIG) at the low ionization regime as seen in \citet{2020ApJ...902L..16J}.
The strong [O~\textsc{iii}]$\lambda$5007 emission-line compact galaxies in addition to the local H \textsc{ii} regions are more appropriate local comparison samples for high-$z$ star-forming galaxies on investigating the ionization parameters \citep{2017ApJ...850..136S}.
The offset might be explained by harder ionizing sources at higher redshifts \citep{2017ApJ...836..164S,2019ApJ...881L..35S} or the evolution of the fundamental metallicity relation \citep{2016ApJ...816...23S}. 

\begin{figure*}[!ht]
  \centering
  \includegraphics[width=\hsize]{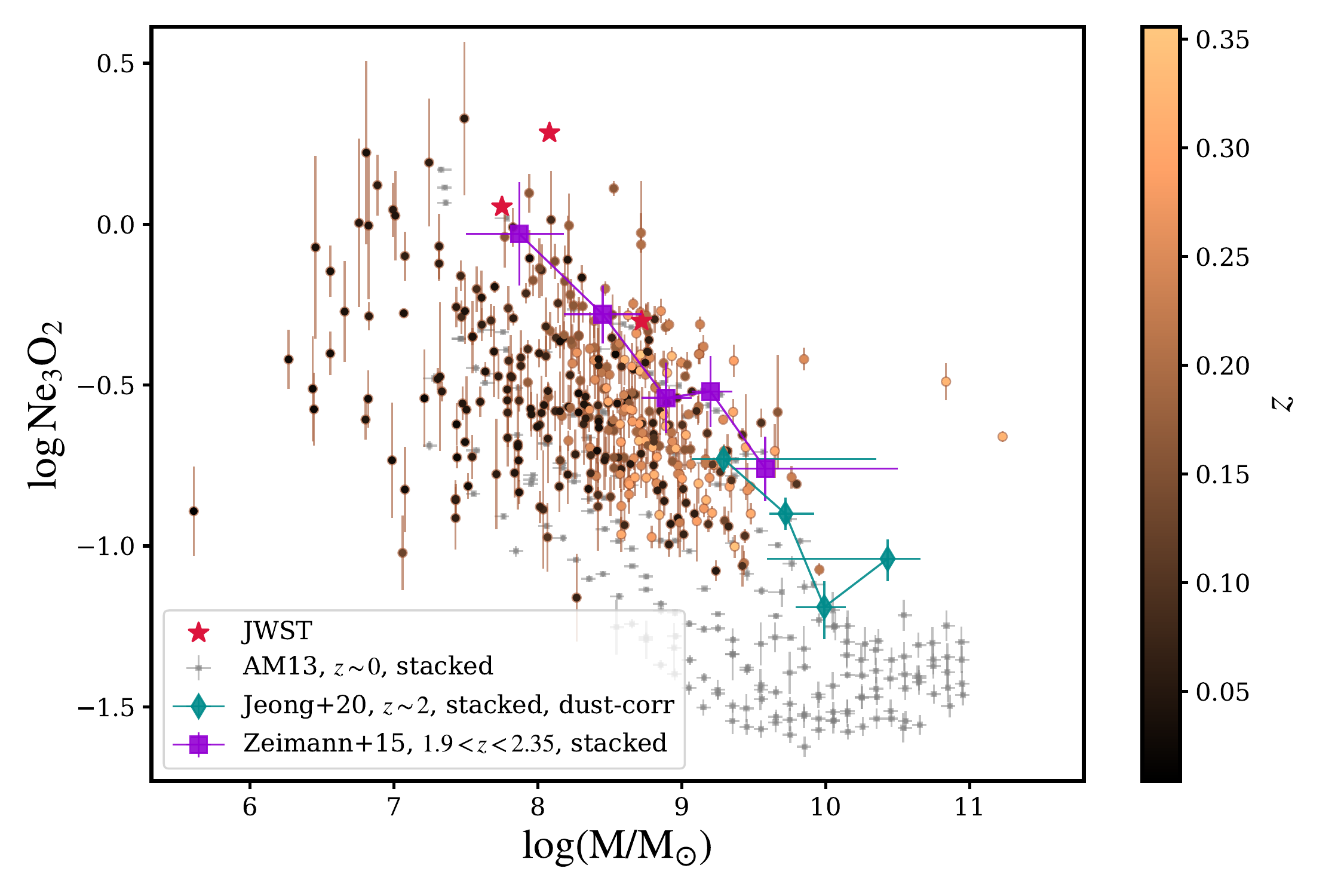}
      \caption{The relation of the Ne3O2 vs stellar mass of the strong [O~\textsc{iii}]$\lambda$5007 emission-line compact galaxies, where the colors of the scatter points mark the redshift of the source. 
      The samples at $z\sim 0$ from \citet{2013ApJ...765..140A} are marked with gray stars, the stacked measurement results with the dust corrected from \citet{2020ApJ...902L..16J} are marked with dark cyan diamonds, the stacked measurement results from the \citet{2020ApJ...902L..16J} are marked with dark orchid squares \citep{2015ApJ...798...29Z}.
      The crimson stars mark the three galaxies from \textit{JWST}.
      }
         \label{fig:Ne3O2-mass}
  \end{figure*}

In Figure \ref{fig:Ne3O2-mass}, we demonstrate the anti-correlation of the Ne3O2 with the stellar mass at the mass range $\rm \log M/M_{\odot} < 9.5$.
This anti-correlation is also consistent with the mass-metallicity relation, based on the anti-correlation between the ionization parameter and metallicity. 
This trend is also demonstrated in \citet{2023arXiv230107444P} for SDSS galaxies at $z\sim 0$\citep{2013ApJ...765..140A}, and HALO7D galaxies at $z\sim 0.8$\citep{2019ApJ...876..124C,2019ApJ...879..120C}.
At fixed stellar mass, with the increase of redshift, the Ne3O2 is higher for the strong [O~\textsc{iii}]$\lambda$5007 emission-line compact galaxies in this work as demonstrated by the color of the scatter points. 
The Ne3O2 offset is also obvious with measurements from the stacked spectra at $z\sim2$ \citep{2015ApJ...798...29Z,2020ApJ...902L..16J}. 
The small offset between \citet{2015ApJ...798...29Z} and \citet{2020ApJ...902L..16J} is due to the blended He \textsc{i} and H $\zeta$ lines in the low-resolution \textit{HST} grism spectra. 

\subsection{Ionized gas kinematics from the asymmetric [O~\textsc{iii}]$\lambda$5007 profile}
\label{sec:kinematics}
Visually inspecting the [O~\textsc{iii}]$\lambda5007$ emission-line lines, we notice that some sources show significant asymmetric profiles.
By comparing the fitting result of the double-Gaussian model with the single-Gaussian model, we select the model where the $\chi^2_{\nu}$ is closer to 1. 
The degree of freedom is the number of spectral points within the wavelength range of  4985--5029 \r{A} minus 6 for a double-Gaussian model, or minus 3 for a single-Gaussian model. 
We have identified eight sources that show asymmetric [O~\textsc{iii}]$\lambda5007$ emission-line profiles as demonstrated in Figure \ref{fig:asymmetry_1}, where the $\rm \chi^2_{\nu, double}$ is less than 90\% of the $\rm \chi^2_{\nu, single}$. 
The properties for these sources are listed in Table \ref{tab:table_asymmetry}.

To ensure the validness of the asymmetric emission line profiles, we visually check the positions of the bad pixels to avoid contaminating the spectra.
In the original LAMOST spectra, the primary data arrays store the \texttt{andmask} and the \texttt{ormask}\footnote{\url{http://www.lamost.org/dr9/v1.0/doc/lr-data-production-description}}.
The \texttt{andmask} is a decimal integer.  If one of the six situations (bad pixel on CCD, bad profile in extraction, no sky information at this wavelength, sky level too high, fiber trace out of the CCD, or no good data) always appears in each exposure, the \texttt{andmask} will be set to 1. 
Similarly, the \texttt{ormask} is set to 1 if one of the above six situations happens in any of the exposures.
We have checked carefully for the masks in these sources, and none of the H$\beta$, [O~\textsc{iii}]$\lambda5007$ and [O~\textsc{iii}]$\lambda5007$ spectral regions is marked with 1 with either \texttt{allmask} or \texttt{ormask}.

We also check the single-exposure spectra to validate the accuracy of the asymmetry profiles. 
The measured velocity dispersion, the number of exposures, the number of exposures used, and the measurement error are in Table \ref{tab:single_exposure}.
Specifically for source 811511143, we demonstrate double-Gaussian fitting result for both the [O~\textsc{iii}]$\lambda4959$ and [O~\textsc{iii}]$\lambda5007$ emission lines for each of the selected single exposures in Figure \ref{fig:811511143}.

\citet{2012ApJ...749..185A,2012ApJ...754L..22A,2019MNRAS.489.1787B,2020MNRAS.494.3541H,2023arXiv230408284L} have discussed the phenomenon of the double-peak or asymmetric profile in their spectra of the Green Pea galaxies.  
\citet{2012ApJ...754L..22A} have identified multiple components with spatially-resolved offsets in their two-dimensional spectra. 
By analyzing the chemical abundances of each component, \citet{2020MNRAS.494.3541H} conclude that the outflows and turbulence are the driving force for the escape of ionizing photons and chemical enrichment.  
These studies show that the multiple gas component is not unique in Green Pea galaxies but also exists in $z\sim1-2$ extreme emission-line galaxies \citep{2014ApJ...791...17M,2014ApJ...785..153M,2015MNRAS.451.3001T} with high-velocity dispersions up to about 240 km s$^{-1}$, as well as detected in the integral field spectroscopy of other compact star-forming galaxies \citep{2009ApJ...697.2057L,2017ApJ...838....4Y,2017MNRAS.471.1280T} and giant H \textsc{ii} regions \citep{1990ApJ...365..164C}.

Due to the limits of the instrumentation resolution ($R\approx 1800$), it is impossible for us to identify narrower components.
It is worthwhile to conduct follow-up high-resolution spectroscopy with a longer exposure time to confirm the existence of the multi-components.

\section{Conclusions}
\label{sec:conclusions}

Our work aims to learn about the chemical composition and kinematics of the strong [O~\textsc{iii}]$\lambda$5007 emission-line compact galaxies in the LAMOST survey. 
With the photometric catalog of SDSS and spectra database from LAMOST, we newly identify 402 spectra and 346 new strong [O~\textsc{iii}]$\lambda$5007 emission-line compact galaxies by finding compact isolated point sources, and confirm by LAMOST spectra.  
Combined with the Blueberry, Green Pea, and Purple Grape samples in our previous work \citep{2022ApJ...927...57L}, this returns a sample of strong [O~\textsc{iii}]$\lambda$5007 emission-line compact galaxies with 1830 unique strong [O~\textsc{iii}]$\lambda$5007 emission-line compact galaxies with 2033 spectra up to redshift of 0.53.
Using the emission lines, we check the metallicity of these sources. 

Our conclusions are as follows:
\begin{itemize}
\item The strong-line metallicity diagnostics calibrated from the direct-$T_e$ method also apply to the strong [O~\textsc{iii}]$\lambda$5007 emission-line compact galaxies. 
\item The strong [O~\textsc{iii}]$\lambda$5007 emission-line compact galaxies fall below several $T_e$-calibrated mass-metallicity relations, especially \citet{2020A&A...634A.107Y,2020MNRAS.491..944C}.  
\item The N/O vs O/H relation of the strong [O~\textsc{iii}]$\lambda$5007 emission-line compact galaxies mainly follows the relation at low metallicity when the nitrogen is primarily produced from primary nucleosynthesis. 
\item The ionization parameter mapped from Ne3O2 and O32 follows a tight linear relation.  
No significant redshift evolution is observed.  
\item The Ne3O2 anti-correlates with the stellar mass for these strong [O~\textsc{iii}]$\lambda$5007 emission-line compact galaxies at $\rm \log (M_{\star}/M_\odot)< 9.5$.  At fixed stellar mass, the Ne3O2 is higher at higher redshifts. 
\item Three sources from \textit{JWST} follow a similar relation with the strong [O~\textsc{iii}]$\lambda$5007 emission-line compact galaxies in this work in MZR, N/O vs O/H relation, and the Ne3O2 vs O32 relation. 
\item Eight sources with asymmetric [O~\textsc{iii}]$\lambda5007$ emission-line profiles have been identified however with no [O~\textsc{iii}]$\lambda4363$ detection, which proves the rich metal content and complex ionized gas kinematics within the galaxies \citep{2012ApJ...749..185A,2012ApJ...754L..22A,2019MNRAS.489.1787B,2020MNRAS.494.3541H}.  
 However, due to limited spectral resolution, we cannot identify the components with velocity dispersion less than 70 km s$^{-1}$.
\end{itemize}

\newpage
\section*{Acknowledgements}
This work is supported by the National Science Foundation of China (No. 12273075) and the National Key R\&D Program of China (No. 2019YFA0405502).  
Guoshoujing Telescope (the Large Sky Area Multi-Object Fiber Spectroscopic Telescope, LAMOST) is a National Major Scientific Project built by the Chinese Academy of Sciences.  
Funding for the project has been provided by the National Development and Reform Commission.  
LAMOST is operated and managed by the National Astronomical Observatories, the Chinese Academy of Sciences.
W. Zhang acknowledges support from the National Science Foundation of China (No. 12090041),  the National Key R\&D Program of China (No. 2021YFA1600401, 2021YFA1600400), and the Guangxi Natural Science Foundation (No. 2019GXNSFFA245008). 
S.L. thanks for the useful discussion with Jun-Qiang Ge, Qing Liu, Xiang-Lei Chen, and Yun-Jin Zhang.

\facilities{LAMOST, SDSS, WISE, GALEX}
\software{astropy \citep{2013A&A...558A..33A,2018AJ....156..123A},  
          extinction \citep{barbary_kyle_2016_804967},
          LMFIT \citep{2016ascl.soft06014N}, 
          Matplotlib \citep{Hunter:2007},
          Numpy \citep{2020Natur.585..357H}, 
          pandas \citep{mckinney-proc-scipy-2010,reback2020pandas}, 
          Scipy \citep{2020NatMe..17..261V},           
          Source Extractor \citep{1996A&AS..117..393B},
          TOPCAT \citep{2005ASPC..347...29T}.
          }
\newpage
\bibliographystyle{aasjournal}
\bibliography{GP_new.bib}

\begin{thebibliography}{}
\expandafter\ifx\csname natexlab\endcsname\relax\def\natexlab#1{#1}\fi
\providecommand{\url}[1]{\href{#1}{#1}}
\providecommand{\dodoi}[1]{doi:~\href{http://doi.org/#1}{\nolinkurl{#1}}}
\providecommand{\doeprint}[1]{\href{http://ascl.net/#1}{\nolinkurl{http://ascl.net/#1}}}
\providecommand{\doarXiv}[1]{\href{https://arxiv.org/abs/#1}{\nolinkurl{https://arxiv.org/abs/#1}}}

\bibitem[{{Aller}(1984)}]{1984ASSL..112.....A}
{Aller}, L.~H. 1984, {Physics of thermal gaseous nebulae},
  \dodoi{10.1007/978-94-010-9639-3}

\bibitem[{{Alloin} {et~al.}(1979){Alloin}, {Collin-Souffrin}, {Joly}, \&
  {Vigroux}}]{1979A&A....78..200A}
{Alloin}, D., {Collin-Souffrin}, S., {Joly}, M., \& {Vigroux}, L. 1979, \aap,
  78, 200

\bibitem[{{Amor{\'\i}n} {et~al.}(2012{\natexlab{a}}){Amor{\'\i}n},
  {P{\'e}rez-Montero}, {V{\'\i}lchez}, \& {Papaderos}}]{2012ApJ...749..185A}
{Amor{\'\i}n}, R., {P{\'e}rez-Montero}, E., {V{\'\i}lchez}, J.~M., \&
  {Papaderos}, P. 2012{\natexlab{a}}, \apj, 749, 185,
  \dodoi{10.1088/0004-637X/749/2/185}

\bibitem[{{Amor{\'\i}n} {et~al.}(2012{\natexlab{b}}){Amor{\'\i}n},
  {V{\'\i}lchez}, {H{\"a}gele}, {Firpo}, {P{\'e}rez-Montero}, \&
  {Papaderos}}]{2012ApJ...754L..22A}
{Amor{\'\i}n}, R., {V{\'\i}lchez}, J.~M., {H{\"a}gele}, G.~F., {et~al.}
  2012{\natexlab{b}}, \apjl, 754, L22, \dodoi{10.1088/2041-8205/754/2/L22}

\bibitem[{{Amor{\'\i}n} {et~al.}(2010){Amor{\'\i}n}, {P{\'e}rez-Montero}, \&
  {V{\'\i}lchez}}]{2010ApJ...715L.128A}
{Amor{\'\i}n}, R.~O., {P{\'e}rez-Montero}, E., \& {V{\'\i}lchez}, J.~M. 2010,
  \apjl, 715, L128, \dodoi{10.1088/2041-8205/715/2/L128}

\bibitem[{{Andrews} \& {Martini}(2013)}]{2013ApJ...765..140A}
{Andrews}, B.~H., \& {Martini}, P. 2013, \apj, 765, 140,
  \dodoi{10.1088/0004-637X/765/2/140}

\bibitem[{{Astropy Collaboration} {et~al.}(2013){Astropy Collaboration},
  {Robitaille}, {Tollerud}, {Greenfield}, {Droettboom}, {Bray}, {Aldcroft},
  {Davis}, {Ginsburg}, {Price-Whelan}, {Kerzendorf}, {Conley}, {Crighton},
  {Barbary}, {Muna}, {Ferguson}, {Grollier}, {Parikh}, {Nair}, {Unther},
  {Deil}, {Woillez}, {Conseil}, {Kramer}, {Turner}, {Singer}, {Fox}, {Weaver},
  {Zabalza}, {Edwards}, {Azalee Bostroem}, {Burke}, {Casey}, {Crawford},
  {Dencheva}, {Ely}, {Jenness}, {Labrie}, {Lim}, {Pierfederici}, {Pontzen},
  {Ptak}, {Refsdal}, {Servillat}, \& {Streicher}}]{2013A&A...558A..33A}
{Astropy Collaboration}, {Robitaille}, T.~P., {Tollerud}, E.~J., {et~al.} 2013,
  \aap, 558, A33, \dodoi{10.1051/0004-6361/201322068}

\bibitem[{{Astropy Collaboration} {et~al.}(2018){Astropy Collaboration},
  {Price-Whelan}, {Sip{\H{o}}cz}, {G{\"u}nther}, {Lim}, {Crawford}, {Conseil},
  {Shupe}, {Craig}, {Dencheva}, {Ginsburg}, {VanderPlas}, {Bradley},
  {P{\'e}rez-Su{\'a}rez}, {de Val-Borro}, {Aldcroft}, {Cruz}, {Robitaille},
  {Tollerud}, {Ardelean}, {Babej}, {Bach}, {Bachetti}, {Bakanov}, {Bamford},
  {Barentsen}, {Barmby}, {Baumbach}, {Berry}, {Biscani}, {Boquien}, {Bostroem},
  {Bouma}, {Brammer}, {Bray}, {Breytenbach}, {Buddelmeijer}, {Burke},
  {Calderone}, {Cano Rodr{\'\i}guez}, {Cara}, {Cardoso}, {Cheedella}, {Copin},
  {Corrales}, {Crichton}, {D'Avella}, {Deil}, {Depagne}, {Dietrich}, {Donath},
  {Droettboom}, {Earl}, {Erben}, {Fabbro}, {Ferreira}, {Finethy}, {Fox},
  {Garrison}, {Gibbons}, {Goldstein}, {Gommers}, {Greco}, {Greenfield},
  {Groener}, {Grollier}, {Hagen}, {Hirst}, {Homeier}, {Horton}, {Hosseinzadeh},
  {Hu}, {Hunkeler}, {Ivezi{\'c}}, {Jain}, {Jenness}, {Kanarek}, {Kendrew},
  {Kern}, {Kerzendorf}, {Khvalko}, {King}, {Kirkby}, {Kulkarni}, {Kumar},
  {Lee}, {Lenz}, {Littlefair}, {Ma}, {Macleod}, {Mastropietro}, {McCully},
  {Montagnac}, {Morris}, {Mueller}, {Mumford}, {Muna}, {Murphy}, {Nelson},
  {Nguyen}, {Ninan}, {N{\"o}the}, {Ogaz}, {Oh}, {Parejko}, {Parley}, {Pascual},
  {Patil}, {Patil}, {Plunkett}, {Prochaska}, {Rastogi}, {Reddy Janga},
  {Sabater}, {Sakurikar}, {Seifert}, {Sherbert}, {Sherwood-Taylor}, {Shih},
  {Sick}, {Silbiger}, {Singanamalla}, {Singer}, {Sladen}, {Sooley},
  {Sornarajah}, {Streicher}, {Teuben}, {Thomas}, {Tremblay}, {Turner},
  {Terr{\'o}n}, {van Kerkwijk}, {de la Vega}, {Watkins}, {Weaver}, {Whitmore},
  {Woillez}, {Zabalza}, \& {Astropy Contributors}}]{2018AJ....156..123A}
{Astropy Collaboration}, {Price-Whelan}, A.~M., {Sip{\H{o}}cz}, B.~M., {et~al.}
  2018, \aj, 156, 123, \dodoi{10.3847/1538-3881/aabc4f}

\bibitem[{Barbary(2016)}]{barbary_kyle_2016_804967}
Barbary, K. 2016, extinction v0.3.0,  Zenodo, \dodoi{10.5281/zenodo.804967}

\bibitem[{{Bertin} \& {Arnouts}(1996)}]{1996A&AS..117..393B}
{Bertin}, E., \& {Arnouts}, S. 1996, \aaps, 117, 393,
  \dodoi{10.1051/aas:1996164}

\bibitem[{Bianchi {et~al.}(2017)Bianchi, Shiao, \& Thilker}]{Bianchi_2017}
Bianchi, L., Shiao, B., \& Thilker, D. 2017, The Astrophysical Journal
  Supplement Series, 230, 24, \dodoi{10.3847/1538-4365/aa7053}

\bibitem[{{Boquien} {et~al.}(2019){Boquien}, {Burgarella}, {Roehlly}, {Buat},
  {Ciesla}, {Corre}, {Inoue}, \& {Salas}}]{2019A&A...622A.103B}
{Boquien}, M., {Burgarella}, D., {Roehlly}, Y., {et~al.} 2019, \aap, 622, A103,
  \dodoi{10.1051/0004-6361/201834156}

\bibitem[{{Bosch} {et~al.}(2019){Bosch}, {H{\"a}gele}, {Amor{\'\i}n}, {Firpo},
  {Cardaci}, {V{\'\i}lchez}, {P{\'e}rez-Montero}, {Papaderos}, {Dors},
  {Krabbe}, \& {Campuzano-Castro}}]{2019MNRAS.489.1787B}
{Bosch}, G., {H{\"a}gele}, G.~F., {Amor{\'\i}n}, R., {et~al.} 2019, \mnras,
  489, 1787, \dodoi{10.1093/mnras/stz2230}

\bibitem[{Bruzual \& Charlot(2003)}]{bruzual2003stellar}
Bruzual, G., \& Charlot, S. 2003, Monthly Notices of the Royal Astronomical
  Society, 344, 1000

\bibitem[{{Bruzual} \& {Charlot}(2003)}]{2003MNRAS.344.1000B}
{Bruzual}, G., \& {Charlot}, S. 2003, \mnras, 344, 1000,
  \dodoi{10.1046/j.1365-8711.2003.06897.x}

\bibitem[{{Burgarella} {et~al.}(2005){Burgarella}, {Buat}, \&
  {Iglesias-P{\'a}ramo}}]{2005MNRAS.360.1413B}
{Burgarella}, D., {Buat}, V., \& {Iglesias-P{\'a}ramo}, J. 2005, \mnras, 360,
  1413, \dodoi{10.1111/j.1365-2966.2005.09131.x}

\bibitem[{Calzetti {et~al.}(2000)Calzetti, Armus, Bohlin, Kinney, Koornneef, \&
  Storchi-Bergmann}]{calzetti2000dust}
Calzetti, D., Armus, L., Bohlin, R.~C., {et~al.} 2000, The Astrophysical
  Journal, 533, 682

\bibitem[{{Cardamone} {et~al.}(2009){Cardamone}, {Schawinski}, {Sarzi},
  {Bamford}, {Bennert}, {Urry}, {Lintott}, {Keel}, {Parejko}, {Nichol},
  {Thomas}, {Andreescu}, {Murray}, {Raddick}, {Slosar}, {Szalay}, \&
  {Vandenberg}}]{2009MNRAS.399.1191C}
{Cardamone}, C., {Schawinski}, K., {Sarzi}, M., {et~al.} 2009, \mnras, 399,
  1191, \dodoi{10.1111/j.1365-2966.2009.15383.x}

\bibitem[{{Casey}(2012)}]{2012MNRAS.425.3094C}
{Casey}, C.~M. 2012, \mnras, 425, 3094,
  \dodoi{10.1111/j.1365-2966.2012.21455.x}

\bibitem[{{Castaneda} {et~al.}(1990){Castaneda}, {Vilchez}, \&
  {Copetti}}]{1990ApJ...365..164C}
{Castaneda}, H.~O., {Vilchez}, J.~M., \& {Copetti}, M. V.~F. 1990, \apj, 365,
  164, \dodoi{10.1086/169466}

\bibitem[{{Chabrier}(2003)}]{2003PASP..115..763C}
{Chabrier}, G. 2003, \pasp, 115, 763, \dodoi{10.1086/376392}

\bibitem[{{Chiappini} {et~al.}(2005){Chiappini}, {Matteucci}, \&
  {Ballero}}]{2005A&A...437..429C}
{Chiappini}, C., {Matteucci}, F., \& {Ballero}, S.~K. 2005, \aap, 437, 429,
  \dodoi{10.1051/0004-6361:20042292}

\bibitem[{{Cunningham} {et~al.}(2019{\natexlab{a}}){Cunningham}, {Deason},
  {Rockosi}, {Guhathakurta}, {Jennings}, {Kirby}, {Toloba}, \&
  {Barro}}]{2019ApJ...876..124C}
{Cunningham}, E.~C., {Deason}, A.~J., {Rockosi}, C.~M., {et~al.}
  2019{\natexlab{a}}, \apj, 876, 124, \dodoi{10.3847/1538-4357/ab16cb}

\bibitem[{{Cunningham} {et~al.}(2019{\natexlab{b}}){Cunningham}, {Deason},
  {Sanderson}, {Sohn}, {Anderson}, {Guhathakurta}, {Rockosi}, {van der Marel},
  {Loebman}, \& {Wetzel}}]{2019ApJ...879..120C}
{Cunningham}, E.~C., {Deason}, A.~J., {Sanderson}, R.~E., {et~al.}
  2019{\natexlab{b}}, \apj, 879, 120, \dodoi{10.3847/1538-4357/ab24cd}

\bibitem[{{Curti} {et~al.}(2017){Curti}, {Cresci}, {Mannucci}, {Marconi},
  {Maiolino}, \& {Esposito}}]{2017MNRAS.465.1384C}
{Curti}, M., {Cresci}, G., {Mannucci}, F., {et~al.} 2017, \mnras, 465, 1384,
  \dodoi{10.1093/mnras/stw2766}

\bibitem[{{Curti} {et~al.}(2020){Curti}, {Mannucci}, {Cresci}, \&
  {Maiolino}}]{2020MNRAS.491..944C}
{Curti}, M., {Mannucci}, F., {Cresci}, G., \& {Maiolino}, R. 2020, \mnras, 491,
  944, \dodoi{10.1093/mnras/stz2910}

\bibitem[{{Curti} {et~al.}(2023){Curti}, {D'Eugenio}, {Carniani}, {Maiolino},
  {Sandles}, {Witstok}, {Baker}, {Bennett}, {Piotrowska}, {Tacchella},
  {Charlot}, {Nakajima}, {Maheson}, {Mannucci}, {Amiri}, {Arribas}, {Belfiore},
  {Bonaventura}, {Bunker}, {Chevallard}, {Cresci}, {Curtis-Lake},
  {Hayden-Pawson}, {Jones}, {Kumari}, {Laseter}, {Looser}, {Marconi}, {Maseda},
  {Scholtz}, {Smit}, {{\"U}bler}, \& {Wallace}}]{2023MNRAS.518..425C}
{Curti}, M., {D'Eugenio}, F., {Carniani}, S., {et~al.} 2023, \mnras, 518, 425,
  \dodoi{10.1093/mnras/stac2737}

\bibitem[{{Ebeling} {et~al.}(2007){Ebeling}, {Barrett}, {Donovan}, {Ma},
  {Edge}, \& {van Speybroeck}}]{2007ApJ...661L..33E}
{Ebeling}, H., {Barrett}, E., {Donovan}, D., {et~al.} 2007, \apjl, 661, L33,
  \dodoi{10.1086/518603}

\bibitem[{{Ebeling} {et~al.}(2001){Ebeling}, {Edge}, \&
  {Henry}}]{2001ApJ...553..668E}
{Ebeling}, H., {Edge}, A.~C., \& {Henry}, J.~P. 2001, \apj, 553, 668,
  \dodoi{10.1086/320958}

\bibitem[{{Ebeling} {et~al.}(2010){Ebeling}, {Edge}, {Mantz}, {Barrett},
  {Henry}, {Ma}, \& {van Speybroeck}}]{2010MNRAS.407...83E}
{Ebeling}, H., {Edge}, A.~C., {Mantz}, A., {et~al.} 2010, \mnras, 407, 83,
  \dodoi{10.1111/j.1365-2966.2010.16920.x}

\bibitem[{{Ferruit} {et~al.}(2022){Ferruit}, {Jakobsen}, {Giardino}, {Rawle},
  {Alves de Oliveira}, {Arribas}, {Beck}, {Birkmann}, {B{\"o}ker}, {Bunker},
  {Charlot}, {de Marchi}, {Franx}, {Henry}, {Karakla}, {Kassin}, {Kumari},
  {L{\'o}pez-Caniego}, {L{\"u}tzgendorf}, {Maiolino}, {Manjavacas}, {Marston},
  {Moseley}, {Muzerolle}, {Pirzkal}, {Rauscher}, {Rix}, {Sabbi}, {Sirianni},
  {te Plate}, {Valenti}, {Willott}, \& {Zeidler}}]{2022A&A...661A..81F}
{Ferruit}, P., {Jakobsen}, P., {Giardino}, G., {et~al.} 2022, \aap, 661, A81,
  \dodoi{10.1051/0004-6361/202142673}

\bibitem[{{Fritz} {et~al.}(2006){Fritz}, {Franceschini}, \&
  {Hatziminaoglou}}]{2006MNRAS.366..767F}
{Fritz}, J., {Franceschini}, A., \& {Hatziminaoglou}, E. 2006, \mnras, 366,
  767, \dodoi{10.1111/j.1365-2966.2006.09866.x}

\bibitem[{{Fukugita} {et~al.}(1996){Fukugita}, {Ichikawa}, {Gunn}, {Doi},
  {Shimasaku}, \& {Schneider}}]{1996AJ....111.1748F}
{Fukugita}, M., {Ichikawa}, T., {Gunn}, J.~E., {et~al.} 1996, \aj, 111, 1748,
  \dodoi{10.1086/117915}

\bibitem[{{Harris} {et~al.}(2020){Harris}, {Millman}, {van der Walt},
  {Gommers}, {Virtanen}, {Cournapeau}, {Wieser}, {Taylor}, {Berg}, {Smith},
  {Kern}, {Picus}, {Hoyer}, {van Kerkwijk}, {Brett}, {Haldane}, {del R{\'\i}o},
  {Wiebe}, {Peterson}, {G{\'e}rard-Marchant}, {Sheppard}, {Reddy}, {Weckesser},
  {Abbasi}, {Gohlke}, \& {Oliphant}}]{2020Natur.585..357H}
{Harris}, C.~R., {Millman}, K.~J., {van der Walt}, S.~J., {et~al.} 2020, \nat,
  585, 357, \dodoi{10.1038/s41586-020-2649-2}

\bibitem[{{Hayden-Pawson} {et~al.}(2022){Hayden-Pawson}, {Curti}, {Maiolino},
  {Cirasuolo}, {Belfiore}, {Cappellari}, {Concas}, {Cresci}, {Cullen},
  {Kobayashi}, {Mannucci}, {Marconi}, {Meneghetti}, {Mercurio}, {Peng},
  {Swinbank}, \& {Vincenzo}}]{2022MNRAS.512.2867H}
{Hayden-Pawson}, C., {Curti}, M., {Maiolino}, R., {et~al.} 2022, \mnras, 512,
  2867, \dodoi{10.1093/mnras/stac584}

\bibitem[{{Henry} {et~al.}(2015){Henry}, {Scarlata}, {Martin}, \&
  {Erb}}]{2015ApJ...809...19H}
{Henry}, A., {Scarlata}, C., {Martin}, C.~L., \& {Erb}, D. 2015, \apj, 809, 19,
  \dodoi{10.1088/0004-637X/809/1/19}

\bibitem[{{Hinshaw} {et~al.}(2013){Hinshaw}, {Larson}, {Komatsu}, {Spergel},
  {Bennett}, {Dunkley}, {Nolta}, {Halpern}, {Hill}, {Odegard}, {Page}, {Smith},
  {Weiland}, {Gold}, {Jarosik}, {Kogut}, {Limon}, {Meyer}, {Tucker}, {Wollack},
  \& {Wright}}]{2013ApJS..208...19H}
{Hinshaw}, G., {Larson}, D., {Komatsu}, E., {et~al.} 2013, \apjs, 208, 19,
  \dodoi{10.1088/0067-0049/208/2/19}

\bibitem[{{Hogarth} {et~al.}(2020){Hogarth}, {Amor{\'\i}n}, {V{\'\i}lchez},
  {H{\"a}gele}, {Cardaci}, {P{\'e}rez-Montero}, {Firpo}, {Jaskot}, \&
  {Ch{\'a}vez}}]{2020MNRAS.494.3541H}
{Hogarth}, L., {Amor{\'\i}n}, R., {V{\'\i}lchez}, J.~M., {et~al.} 2020, \mnras,
  494, 3541, \dodoi{10.1093/mnras/staa851}

\bibitem[{Hunter(2007)}]{Hunter:2007}
Hunter, J.~D. 2007, Computing in Science \& Engineering, 9, 90,
  \dodoi{10.1109/MCSE.2007.55}

\bibitem[{{Izotov} {et~al.}(2018){Izotov}, {Schaerer}, {Worseck}, {Guseva},
  {Thuan}, {Verhamme}, {Orlitov{\'a}}, \& {Fricke}}]{2018MNRAS.474.4514I}
{Izotov}, Y.~I., {Schaerer}, D., {Worseck}, G., {et~al.} 2018, \mnras, 474,
  4514, \dodoi{10.1093/mnras/stx3115}

\bibitem[{{Izotov} {et~al.}(2006){Izotov}, {Stasi{\'n}ska}, {Meynet}, {Guseva},
  \& {Thuan}}]{2006A&A...448..955I}
{Izotov}, Y.~I., {Stasi{\'n}ska}, G., {Meynet}, G., {Guseva}, N.~G., \&
  {Thuan}, T.~X. 2006, \aap, 448, 955, \dodoi{10.1051/0004-6361:20053763}

\bibitem[{{Jakobsen} {et~al.}(2022){Jakobsen}, {Ferruit}, {Alves de Oliveira},
  {Arribas}, {Bagnasco}, {Barho}, {Beck}, {Birkmann}, {B{\"o}ker}, {Bunker},
  {Charlot}, {de Jong}, {de Marchi}, {Ehrenwinkler}, {Falcolini}, {Fels},
  {Franx}, {Franz}, {Funke}, {Giardino}, {Gnata}, {Holota}, {Honnen}, {Jensen},
  {Jentsch}, {Johnson}, {Jollet}, {Karl}, {Kling}, {K{\"o}hler}, {Kolm},
  {Kumari}, {Lander}, {Lemke}, {L{\'o}pez-Caniego}, {L{\"u}tzgendorf},
  {Maiolino}, {Manjavacas}, {Marston}, {Maschmann}, {Maurer}, {Messerschmidt},
  {Moseley}, {Mosner}, {Mott}, {Muzerolle}, {Pirzkal}, {Pittet}, {Plitzke},
  {Posselt}, {Rapp}, {Rauscher}, {Rawle}, {Rix}, {R{\"o}del}, {Rumler},
  {Sabbi}, {Salvignol}, {Schmid}, {Sirianni}, {Smith}, {Strada}, {te Plate},
  {Valenti}, {Wettemann}, {Wiehe}, {Wiesmayer}, {Willott}, {Wright}, {Zeidler},
  \& {Zincke}}]{2022A&A...661A..80J}
{Jakobsen}, P., {Ferruit}, P., {Alves de Oliveira}, C., {et~al.} 2022, \aap,
  661, A80, \dodoi{10.1051/0004-6361/202142663}

\bibitem[{{Jaskot} \& {Oey}(2014)}]{2014ApJ...791L..19J}
{Jaskot}, A.~E., \& {Oey}, M.~S. 2014, \apjl, 791, L19,
  \dodoi{10.1088/2041-8205/791/2/L19}

\bibitem[{{Jeong} {et~al.}(2020){Jeong}, {Shapley}, {Sanders}, {Runco},
  {Topping}, {Reddy}, {Kriek}, {Coil}, {Mobasher}, {Siana}, {Shivaei},
  {Freeman}, {Azadi}, {Price}, {Leung}, {Fetherolf}, {de Groot}, {Zick},
  {Fornasini}, \& {Barro}}]{2020ApJ...902L..16J}
{Jeong}, M.-S., {Shapley}, A.~E., {Sanders}, R.~L., {et~al.} 2020, \apjl, 902,
  L16, \dodoi{10.3847/2041-8213/abba7a}

\bibitem[{{Jiang} {et~al.}(2019){Jiang}, {Malhotra}, {Rhoads}, \&
  {Yang}}]{2019ApJ...872..145J}
{Jiang}, T., {Malhotra}, S., {Rhoads}, J.~E., \& {Yang}, H. 2019, \apj, 872,
  145, \dodoi{10.3847/1538-4357/aaee8a}

\bibitem[{{Jones} {et~al.}(2015){Jones}, {Martin}, \&
  {Cooper}}]{2015ApJ...813..126J}
{Jones}, T., {Martin}, C., \& {Cooper}, M.~C. 2015, \apj, 813, 126,
  \dodoi{10.1088/0004-637X/813/2/126}

\bibitem[{{Kewley} \& {Ellison}(2008)}]{2008ApJ...681.1183K}
{Kewley}, L.~J., \& {Ellison}, S.~L. 2008, \apj, 681, 1183,
  \dodoi{10.1086/587500}

\bibitem[{{Kojima} {et~al.}(2020){Kojima}, {Ouchi}, {Rauch}, {Ono}, {Nakajima},
  {Isobe}, {Fujimoto}, {Harikane}, {Hashimoto}, {Hayashi}, {Komiyama},
  {Kusakabe}, {Kim}, {Lee}, {Mukae}, {Nagao}, {Onodera}, {Shibuya}, {Sugahara},
  {Umemura}, \& {Yabe}}]{2020ApJ...898..142K}
{Kojima}, T., {Ouchi}, M., {Rauch}, M., {et~al.} 2020, \apj, 898, 142,
  \dodoi{10.3847/1538-4357/aba047}

\bibitem[{{Komarova} {et~al.}(2021){Komarova}, {Oey}, {Krumholz}, {Silich},
  {Kumari}, \& {James}}]{2021ApJ...920L..46K}
{Komarova}, L., {Oey}, M.~S., {Krumholz}, M.~R., {et~al.} 2021, \apjl, 920,
  L46, \dodoi{10.3847/2041-8213/ac2c09}

\bibitem[{{Kriek} {et~al.}(2015){Kriek}, {Shapley}, {Reddy}, {Siana}, {Coil},
  {Mobasher}, {Freeman}, {de Groot}, {Price}, {Sanders}, {Shivaei}, {Brammer},
  {Momcheva}, {Skelton}, {van Dokkum}, {Whitaker}, {Aird}, {Azadi}, {Kassis},
  {Bullock}, {Conroy}, {Dav{\'e}}, {Kere{\v{s}}}, \&
  {Krumholz}}]{2015ApJS..218...15K}
{Kriek}, M., {Shapley}, A.~E., {Reddy}, N.~A., {et~al.} 2015, \apjs, 218, 15,
  \dodoi{10.1088/0067-0049/218/2/15}

\bibitem[{{Kroupa}(2001)}]{2001MNRAS.322..231K}
{Kroupa}, P. 2001, \mnras, 322, 231, \dodoi{10.1046/j.1365-8711.2001.04022.x}

\bibitem[{{Lara-L{\'o}pez} {et~al.}(2010){Lara-L{\'o}pez}, {Cepa},
  {Bongiovanni}, {P{\'e}rez Garc{\'\i}a}, {Ederoclite}, {Casta{\~n}eda},
  {Fern{\'a}ndez Lorenzo}, {Povi{\'c}}, \&
  {S{\'a}nchez-Portal}}]{2010A&A...521L..53L}
{Lara-L{\'o}pez}, M.~A., {Cepa}, J., {Bongiovanni}, A., {et~al.} 2010, \aap,
  521, L53, \dodoi{10.1051/0004-6361/201014803}

\bibitem[{{Law} {et~al.}(2009){Law}, {Steidel}, {Erb}, {Larkin}, {Pettini},
  {Shapley}, \& {Wright}}]{2009ApJ...697.2057L}
{Law}, D.~R., {Steidel}, C.~C., {Erb}, D.~K., {et~al.} 2009, \apj, 697, 2057,
  \dodoi{10.1088/0004-637X/697/2/2057}

\bibitem[{{Lee} {et~al.}(2006){Lee}, {Skillman}, {Cannon}, {Jackson}, {Gehrz},
  {Polomski}, \& {Woodward}}]{2006ApJ...647..970L}
{Lee}, H., {Skillman}, E.~D., {Cannon}, J.~M., {et~al.} 2006, \apj, 647, 970,
  \dodoi{10.1086/505573}

\bibitem[{{Lequeux} {et~al.}(1979){Lequeux}, {Peimbert}, {Rayo}, {Serrano}, \&
  {Torres-Peimbert}}]{1979A&A....80..155L}
{Lequeux}, J., {Peimbert}, M., {Rayo}, J.~F., {Serrano}, A., \&
  {Torres-Peimbert}, S. 1979, \aap, 80, 155

\bibitem[{{Levesque} \& {Richardson}(2014)}]{2014ApJ...780..100L}
{Levesque}, E.~M., \& {Richardson}, M. L.~A. 2014, \apj, 780, 100,
  \dodoi{10.1088/0004-637X/780/1/100}

\bibitem[{{Lin} {et~al.}(2023){Lin}, {Zheng}, {Wang}, {Yuan}, {Rhoads},
  {Malhotra}, {An}, {Jiang}, {Zhu}, {Rahna P.}, {Ji}, \&
  {Singha}}]{2023arXiv230408284L}
{Lin}, R., {Zheng}, Z.-Y., {Wang}, J.-X., {et~al.} 2023, arXiv e-prints,
  arXiv:2304.08284, \dodoi{10.48550/arXiv.2304.08284}

\bibitem[{{Liu} {et~al.}(2022){Liu}, {Luo}, {Yang}, {Shen}, {Wang}, {Zhang},
  {Zheng}, {Song}, {Kong}, {Wang}, \& {Chen}}]{2022ApJ...927...57L}
{Liu}, S., {Luo}, A.~L., {Yang}, H., {et~al.} 2022, \apj, 927, 57,
  \dodoi{10.3847/1538-4357/ac4bd9}

\bibitem[{{L{\'o}pez-S{\'a}nchez} {et~al.}(2012){L{\'o}pez-S{\'a}nchez},
  {Dopita}, {Kewley}, {Zahid}, {Nicholls}, \&
  {Scharw{\"a}chter}}]{2012MNRAS.426.2630L}
{L{\'o}pez-S{\'a}nchez}, {\'A}.~R., {Dopita}, M.~A., {Kewley}, L.~J., {et~al.}
  2012, \mnras, 426, 2630, \dodoi{10.1111/j.1365-2966.2012.21145.x}

\bibitem[{{Luo} {et~al.}(2012){Luo}, {Zhang}, {Zhao}, {Zhao}, {Cui}, {Li},
  {Chu}, {Shi}, {Wang}, {Zhang}, {Bai}, {Chen}, {Wang}, {Guo}, {Chen}, {Du},
  {Kong}, {Lei}, {Li}, {Song}, {Wu}, {Zhang}, {Zhou}, {Zuo}, {Du}, {He}, {Hou},
  {Dong}, {Li}, {Li}, {Li}, {Song}, {Tian}, {Wang}, {Wu}, {Yang}, {Yuan},
  {Cao}, {Chen}, {Chen}, {Chen}, {Chu}, {Feng}, {Gong}, {Gu}, {Hou}, {Huo},
  {Hu}, {Hu}, {Hu}, {Jia}, {Jiang}, {Jiang}, {Jiang}, {Jin}, {Li}, {Li}, {Li},
  {Li}, {Li}, {Liu}, {Liu}, {Liu}, {Lu}, {Lu}, {Luo}, {Mao}, {Men}, {Ni}, {Qi},
  {Qi}, {Shi}, {Su}, {Sun}, {Su}, {Tang}, {Tao}, {Tu}, {Wang}, {Wang}, {Wang},
  {Wang}, {Wang}, {Wang}, {Wang}, {Wang}, {Wang}, {Wang}, {Wang}, {Wang},
  {Wang}, {Wang}, {Wei}, {Xue}, {Xing}, {Xu}, {Xu}, {Xu}, {Yang}, {Yang},
  {Yao}, {Yu}, {Yuan}, {Zhai}, {Zhang}, {Zhang}, {Zhang}, {Zhang}, {Zhang},
  {Zhang}, {Zhao}, {Zhou}, {Zhu}, {Zhu}, \& {Zou}}]{2012RAA....12.1243L}
{Luo}, A.~L., {Zhang}, H.-T., {Zhao}, Y.-H., {et~al.} 2012, Research in
  Astronomy and Astrophysics, 12, 1243, \dodoi{10.1088/1674-4527/12/9/004}

\bibitem[{{Maiolino} \& {Mannucci}(2019)}]{2019A&ARv..27....3M}
{Maiolino}, R., \& {Mannucci}, F. 2019, \aapr, 27, 3,
  \dodoi{10.1007/s00159-018-0112-2}

\bibitem[{{Maiolino} {et~al.}(2008){Maiolino}, {Nagao}, {Grazian}, {Cocchia},
  {Marconi}, {Mannucci}, {Cimatti}, {Pipino}, {Ballero}, {Calura}, {Chiappini},
  {Fontana}, {Granato}, {Matteucci}, {Pastorini}, {Pentericci}, {Risaliti},
  {Salvati}, \& {Silva}}]{2008A&A...488..463M}
{Maiolino}, R., {Nagao}, T., {Grazian}, A., {et~al.} 2008, \aap, 488, 463,
  \dodoi{10.1051/0004-6361:200809678}

\bibitem[{{Mann} \& {Ebeling}(2012)}]{2012MNRAS.420.2120M}
{Mann}, A.~W., \& {Ebeling}, H. 2012, \mnras, 420, 2120,
  \dodoi{10.1111/j.1365-2966.2011.20170.x}

\bibitem[{{Mannucci} {et~al.}(2010){Mannucci}, {Cresci}, {Maiolino}, {Marconi},
  \& {Gnerucci}}]{2010MNRAS.408.2115M}
{Mannucci}, F., {Cresci}, G., {Maiolino}, R., {Marconi}, A., \& {Gnerucci}, A.
  2010, \mnras, 408, 2115, \dodoi{10.1111/j.1365-2966.2010.17291.x}

\bibitem[{{Maseda} {et~al.}(2014){Maseda}, {van der Wel}, {Rix}, {da Cunha},
  {Pacifici}, {Momcheva}, {Brammer}, {Meidt}, {Franx}, {van Dokkum},
  {Fumagalli}, {Bell}, {Ferguson}, {F{\"o}rster-Schreiber}, {Koekemoer}, {Koo},
  {Lundgren}, {Marchesini}, {Nelson}, {Patel}, {Skelton}, {Straughn}, {Trump},
  \& {Whitaker}}]{2014ApJ...791...17M}
{Maseda}, M.~V., {van der Wel}, A., {Rix}, H.-W., {et~al.} 2014, \apj, 791, 17,
  \dodoi{10.1088/0004-637X/791/1/17}

\bibitem[{{Masters} {et~al.}(2014){Masters}, {McCarthy}, {Siana}, {Malkan},
  {Mobasher}, {Atek}, {Henry}, {Martin}, {Rafelski}, {Hathi}, {Scarlata},
  {Ross}, {Bunker}, {Blanc}, {Bedregal}, {Dom{\'\i}nguez}, {Colbert},
  {Teplitz}, \& {Dressler}}]{2014ApJ...785..153M}
{Masters}, D., {McCarthy}, P., {Siana}, B., {et~al.} 2014, \apj, 785, 153,
  \dodoi{10.1088/0004-637X/785/2/153}

\bibitem[{{Moll{\'a}} {et~al.}(2006){Moll{\'a}}, {V{\'\i}lchez}, {Gavil{\'a}n},
  \& {D{\'\i}az}}]{2006MNRAS.372.1069M}
{Moll{\'a}}, M., {V{\'\i}lchez}, J.~M., {Gavil{\'a}n}, M., \& {D{\'\i}az},
  A.~I. 2006, \mnras, 372, 1069, \dodoi{10.1111/j.1365-2966.2006.10892.x}

\bibitem[{{Nagao} {et~al.}(2006){Nagao}, {Maiolino}, \&
  {Marconi}}]{2006A&A...459...85N}
{Nagao}, T., {Maiolino}, R., \& {Marconi}, A. 2006, \aap, 459, 85,
  \dodoi{10.1051/0004-6361:20065216}

\bibitem[{{Nakajima} {et~al.}(2022){Nakajima}, {Ouchi}, {Xu}, {Rauch},
  {Harikane}, {Nishigaki}, {Isobe}, {Kusakabe}, {Nagao}, {Ono}, {Onodera},
  {Sugahara}, {Kim}, {Komiyama}, {Lee}, \& {Zahedy}}]{2022ApJS..262....3N}
{Nakajima}, K., {Ouchi}, M., {Xu}, Y., {et~al.} 2022, \apjs, 262, 3,
  \dodoi{10.3847/1538-4365/ac7710}

\bibitem[{{Newville} {et~al.}(2016){Newville}, {Stensitzki}, {Allen}, {Rawlik},
  {Ingargiola}, \& {Nelson}}]{2016ascl.soft06014N}
{Newville}, M., {Stensitzki}, T., {Allen}, D.~B., {et~al.} 2016, {Lmfit:
  Non-Linear Least-Square Minimization and Curve-Fitting for Python}.
\newblock \doeprint{1606.014}

\bibitem[{{Oke} \& {Gunn}(1983)}]{1983ApJ...266..713O}
{Oke}, J.~B., \& {Gunn}, J.~E. 1983, \apj, 266, 713, \dodoi{10.1086/160817}

\bibitem[{{Pagel}(1997)}]{1997nceg.book.....P}
{Pagel}, B. E.~J. 1997, {Nucleosynthesis and Chemical Evolution of Galaxies}

\bibitem[{{Pagel} {et~al.}(1979){Pagel}, {Edmunds}, {Blackwell}, {Chun}, \&
  {Smith}}]{1979MNRAS.189...95P}
{Pagel}, B.~E.~J., {Edmunds}, M.~G., {Blackwell}, D.~E., {Chun}, M.~S., \&
  {Smith}, G. 1979, \mnras, 189, 95, \dodoi{10.1093/mnras/189.1.95}

\bibitem[{{Pagel} {et~al.}(1992){Pagel}, {Simonson}, {Terlevich}, \&
  {Edmunds}}]{1992MNRAS.255..325P}
{Pagel}, B.~E.~J., {Simonson}, E.~A., {Terlevich}, R.~J., \& {Edmunds}, M.~G.
  1992, \mnras, 255, 325, \dodoi{10.1093/mnras/255.2.325}

\bibitem[{pandas~development team(2020)}]{reback2020pandas}
pandas~development team, T. 2020, pandas-dev/pandas: Pandas, latest,  Zenodo,
  \dodoi{10.5281/zenodo.3509134}

\bibitem[{{Patr{\'\i}cio} {et~al.}(2018){Patr{\'\i}cio}, {Christensen},
  {Rhodin}, {Ca{\~n}ameras}, \& {Lara-L{\'o}pez}}]{2018MNRAS.481.3520P}
{Patr{\'\i}cio}, V., {Christensen}, L., {Rhodin}, H., {Ca{\~n}ameras}, R., \&
  {Lara-L{\'o}pez}, M.~A. 2018, \mnras, 481, 3520,
  \dodoi{10.1093/mnras/sty2508}

\bibitem[{{P{\'e}rez-Montero} {et~al.}(2007){P{\'e}rez-Montero}, {H{\"a}gele},
  {Contini}, \& {D{\'\i}az}}]{2007MNRAS.381..125P}
{P{\'e}rez-Montero}, E., {H{\"a}gele}, G.~F., {Contini}, T., \& {D{\'\i}az},
  {\'A}.~I. 2007, \mnras, 381, 125, \dodoi{10.1111/j.1365-2966.2007.12213.x}

\bibitem[{{Pharo} {et~al.}(2023){Pharo}, {Guo}, {Koo}, {Forbes}, \&
  {Guhathakurta}}]{2023arXiv230107444P}
{Pharo}, J., {Guo}, Y., {Koo}, D.~C., {Forbes}, J.~C., \& {Guhathakurta}, P.
  2023, arXiv e-prints, arXiv:2301.07444, \dodoi{10.48550/arXiv.2301.07444}

\bibitem[{{Pilyugin} {et~al.}(2012){Pilyugin}, {Grebel}, \&
  {Mattsson}}]{2012MNRAS.424.2316P}
{Pilyugin}, L.~S., {Grebel}, E.~K., \& {Mattsson}, L. 2012, \mnras, 424, 2316,
  \dodoi{10.1111/j.1365-2966.2012.21398.x}

\bibitem[{{Sanders} {et~al.}(2017){Sanders}, {Shapley}, {Zhang}, \&
  {Yan}}]{2017ApJ...850..136S}
{Sanders}, R.~L., {Shapley}, A.~E., {Zhang}, K., \& {Yan}, R. 2017, \apj, 850,
  136, \dodoi{10.3847/1538-4357/aa93e4}

\bibitem[{{Sanders} {et~al.}(2016){Sanders}, {Shapley}, {Kriek}, {Reddy},
  {Freeman}, {Coil}, {Siana}, {Mobasher}, {Shivaei}, {Price}, \& {de
  Groot}}]{2016ApJ...816...23S}
{Sanders}, R.~L., {Shapley}, A.~E., {Kriek}, M., {et~al.} 2016, \apj, 816, 23,
  \dodoi{10.3847/0004-637X/816/1/23}

\bibitem[{{Shapley} {et~al.}(2019){Shapley}, {Sanders}, {Shao}, {Reddy},
  {Kriek}, {Coil}, {Mobasher}, {Siana}, {Shivaei}, {Freeman}, {Azadi}, {Price},
  {Leung}, {Fetherolf}, {de Groot}, {Zick}, {Fornasini}, \&
  {Barro}}]{2019ApJ...881L..35S}
{Shapley}, A.~E., {Sanders}, R.~L., {Shao}, P., {et~al.} 2019, \apjl, 881, L35,
  \dodoi{10.3847/2041-8213/ab385a}

\bibitem[{{Shi} {et~al.}(2005){Shi}, {Kong}, {Li}, \&
  {Cheng}}]{2005A&A...437..849S}
{Shi}, F., {Kong}, X., {Li}, C., \& {Cheng}, F.~Z. 2005, \aap, 437, 849,
  \dodoi{10.1051/0004-6361:20041945}

\bibitem[{{Shi} {et~al.}(2007){Shi}, {Zhao}, \& {Liang}}]{2007A&A...475..409S}
{Shi}, F., {Zhao}, G., \& {Liang}, Y.~C. 2007, \aap, 475, 409,
  \dodoi{10.1051/0004-6361:20077183}

\bibitem[{{Shi} {et~al.}(2014){Shi}, {Luo}, {Comte}, {Chen}, {Wei}, {Zhao},
  {Wu}, {Zhang}, {Shen}, {Yang}, {Wu}, {Wu}, {Zhang}, {Lei}, {Zhang}, {Wang},
  {Jin}, \& {Zhang}}]{2014RAA....14.1234S}
{Shi}, Z.-X., {Luo}, A.~L., {Comte}, G., {et~al.} 2014, Research in Astronomy
  and Astrophysics, 14, 1234, \dodoi{10.1088/1674-4527/14/10/003}

\bibitem[{{Spite} {et~al.}(2005){Spite}, {Cayrel}, {Plez}, {Hill}, {Spite},
  {Depagne}, {Fran{\c{c}}ois}, {Bonifacio}, {Barbuy}, {Beers}, {Andersen},
  {Molaro}, {Nordstr{\"o}m}, \& {Primas}}]{2005A&A...430..655S}
{Spite}, M., {Cayrel}, R., {Plez}, B., {et~al.} 2005, \aap, 430, 655,
  \dodoi{10.1051/0004-6361:20041274}

\bibitem[{{Strom} {et~al.}(2017){Strom}, {Steidel}, {Rudie}, {Trainor},
  {Pettini}, \& {Reddy}}]{2017ApJ...836..164S}
{Strom}, A.~L., {Steidel}, C.~C., {Rudie}, G.~C., {et~al.} 2017, \apj, 836,
  164, \dodoi{10.3847/1538-4357/836/2/164}

\bibitem[{{Su} \& {Cui}(2004)}]{2004ChJAA...4....1S}
{Su}, D.-Q., \& {Cui}, X.-Q. 2004, \cjaa, 4, 1, \dodoi{10.1088/1009-9271/4/1/1}

\bibitem[{{Taylor}(2005)}]{2005ASPC..347...29T}
{Taylor}, M.~B. 2005, in Astronomical Society of the Pacific Conference Series,
  Vol. 347, Astronomical Data Analysis Software and Systems XIV, ed.
  P.~{Shopbell}, M.~{Britton}, \& R.~{Ebert}, 29

\bibitem[{{Terlevich} {et~al.}(2015){Terlevich}, {Terlevich}, {Melnick},
  {Ch{\'a}vez}, {Plionis}, {Bresolin}, \& {Basilakos}}]{2015MNRAS.451.3001T}
{Terlevich}, R., {Terlevich}, E., {Melnick}, J., {et~al.} 2015, \mnras, 451,
  3001, \dodoi{10.1093/mnras/stv1128}

\bibitem[{{Tremonti} {et~al.}(2004){Tremonti}, {Heckman}, {Kauffmann},
  {Brinchmann}, {Charlot}, {White}, {Seibert}, {Peng}, {Schlegel}, {Uomoto},
  {Fukugita}, \& {Brinkmann}}]{2004ApJ...613..898T}
{Tremonti}, C.~A., {Heckman}, T.~M., {Kauffmann}, G., {et~al.} 2004, \apj, 613,
  898, \dodoi{10.1086/423264}

\bibitem[{{Turner} {et~al.}(2017){Turner}, {Cirasuolo}, {Harrison}, {McLure},
  {Dunlop}, {Swinbank}, {Johnson}, {Sobral}, {Matthee}, \&
  {Sharples}}]{2017MNRAS.471.1280T}
{Turner}, O.~J., {Cirasuolo}, M., {Harrison}, C.~M., {et~al.} 2017, \mnras,
  471, 1280, \dodoi{10.1093/mnras/stx1366}

\bibitem[{{Vincenzo} {et~al.}(2016){Vincenzo}, {Belfiore}, {Maiolino},
  {Matteucci}, \& {Ventura}}]{2016MNRAS.458.3466V}
{Vincenzo}, F., {Belfiore}, F., {Maiolino}, R., {Matteucci}, F., \& {Ventura},
  P. 2016, \mnras, 458, 3466, \dodoi{10.1093/mnras/stw532}

\bibitem[{{Virtanen} {et~al.}(2020){Virtanen}, {Gommers}, {Oliphant},
  {Haberland}, {Reddy}, {Cournapeau}, {Burovski}, {Peterson}, {Weckesser},
  {Bright}, {van der Walt}, {Brett}, {Wilson}, {Millman}, {Mayorov}, {Nelson},
  {Jones}, {Kern}, {Larson}, {Carey}, {Polat}, {Feng}, {Moore}, {VanderPlas},
  {Laxalde}, {Perktold}, {Cimrman}, {Henriksen}, {Quintero}, {Harris},
  {Archibald}, {Ribeiro}, {Pedregosa}, {van Mulbregt}, \& {SciPy 1. 0
  Contributors}}]{2020NatMe..17..261V}
{Virtanen}, P., {Gommers}, R., {Oliphant}, T.~E., {et~al.} 2020, Nature
  Methods, 17, 261, \dodoi{10.1038/s41592-019-0686-2}

\bibitem[{{Wang} {et~al.}(2018){Wang}, {Luo}, {Shen}, {Hou}, {Kong}, {Song},
  {Zhang}, {Wu}, {Cao}, {Hou}, {Wang}, {Zhang}, \&
  {Zhao}}]{2018MNRAS.474.1873W}
{Wang}, L.-L., {Luo}, A.~L., {Shen}, S.-Y., {et~al.} 2018, \mnras, 474, 1873,
  \dodoi{10.1093/mnras/stx2798}

\bibitem[{{Wang} {et~al.}(1996){Wang}, {Su}, {Chu}, {Cui}, \&
  {Wang}}]{1996ApOpt..35.5155W}
{Wang}, S.-G., {Su}, D.-Q., {Chu}, Y.-Q., {Cui}, X., \& {Wang}, Y.-N. 1996,
  \ao, 35, 5155, \dodoi{10.1364/AO.35.005155}

\bibitem[{{W}es {M}c{K}inney(2010)}]{mckinney-proc-scipy-2010}
{W}es {M}c{K}inney. 2010, in {P}roceedings of the 9th {P}ython in {S}cience
  {C}onference, ed. {S}t\'efan van~der {W}alt \& {J}arrod {M}illman, 56--61,
  \dodoi{10.25080/Majora-92bf1922-00a}

\bibitem[{{Wright} {et~al.}(2010){Wright}, {Eisenhardt}, {Mainzer}, {Ressler},
  {Cutri}, {Jarrett}, {Kirkpatrick}, {Padgett}, {McMillan}, {Skrutskie},
  {Stanford}, {Cohen}, {Walker}, {Mather}, {Leisawitz}, {Gautier}, {McLean},
  {Benford}, {Lonsdale}, {Blain}, {Mendez}, {Irace}, {Duval}, {Liu}, {Royer},
  {Heinrichsen}, {Howard}, {Shannon}, {Kendall}, {Walsh}, {Larsen}, {Cardon},
  {Schick}, {Schwalm}, {Abid}, {Fabinsky}, {Naes}, \&
  {Tsai}}]{2010AJ....140.1868W}
{Wright}, E.~L., {Eisenhardt}, P. R.~M., {Mainzer}, A.~K., {et~al.} 2010, \aj,
  140, 1868, \dodoi{10.1088/0004-6256/140/6/1868}

\bibitem[{{{Wright}, Edward L. and {Eisenhardt}, Peter R. M. and {Mainzer}, Amy
  K. et al.}(2019)}]{Wright_2019-os}
{{Wright}, Edward L. and {Eisenhardt}, Peter R. M. and {Mainzer}, Amy K. et
  al.} 2019, {{AllWISE} Source Catalog},  IPAC, \dodoi{10.26131/IRSA1}

\bibitem[{{Yang} {et~al.}(2016){Yang}, {Malhotra}, {Gronke}, {Rhoads},
  {Dijkstra}, {Jaskot}, {Zheng}, \& {Wang}}]{2016ApJ...820..130Y}
{Yang}, H., {Malhotra}, S., {Gronke}, M., {et~al.} 2016, \apj, 820, 130,
  \dodoi{10.3847/0004-637X/820/2/130}

\bibitem[{{Yang} {et~al.}(2017){Yang}, {Malhotra}, {Rhoads}, {Leitherer},
  {Wofford}, {Jiang}, \& {Wang}}]{2017ApJ...838....4Y}
{Yang}, H., {Malhotra}, S., {Rhoads}, J.~E., {et~al.} 2017, \apj, 838, 4,
  \dodoi{10.3847/1538-4357/aa6337}

\bibitem[{{Yates} {et~al.}(2020){Yates}, {Schady}, {Chen}, {Schweyer}, \&
  {Wiseman}}]{2020A&A...634A.107Y}
{Yates}, R.~M., {Schady}, P., {Chen}, T.~W., {Schweyer}, T., \& {Wiseman}, P.
  2020, \aap, 634, A107, \dodoi{10.1051/0004-6361/201936506}

\bibitem[{{Zeimann} {et~al.}(2015){Zeimann}, {Ciardullo}, {Gebhardt},
  {Gronwall}, {Hagen}, {Trump}, {Bridge}, {Luo}, \&
  {Schneider}}]{2015ApJ...798...29Z}
{Zeimann}, G.~R., {Ciardullo}, R., {Gebhardt}, H., {et~al.} 2015, \apj, 798,
  29, \dodoi{10.1088/0004-637X/798/1/29}

\end{thebibliography}
\newpage
\appendix

\section{Asymmetric [O~\textsc{iii}]$\lambda5007$ emission-line profiles}

\begin{figure*}[!ht]
  \centering
\subfloat{\includegraphics[width=0.45\hsize]{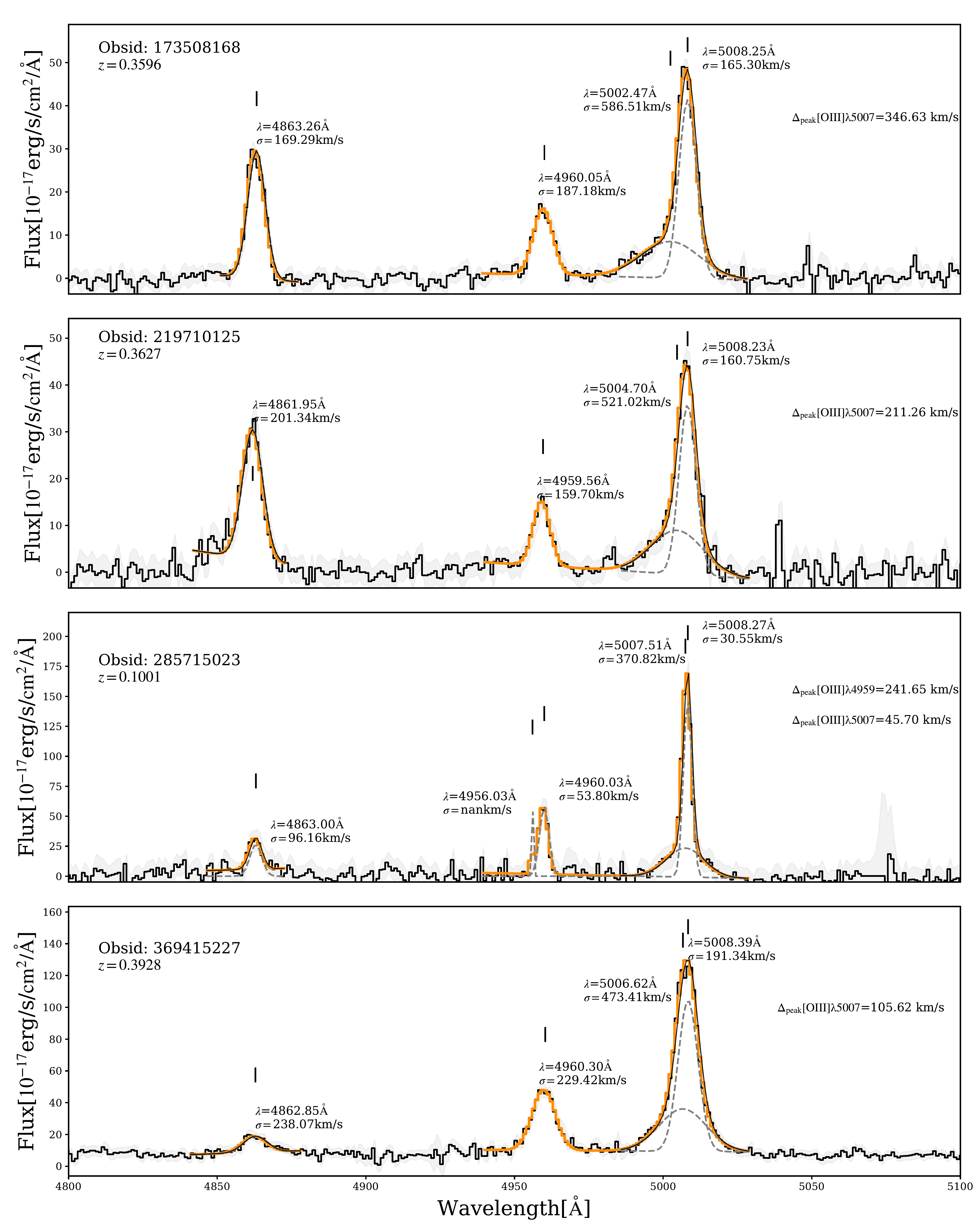}}
\subfloat{\includegraphics[width=0.45\hsize]{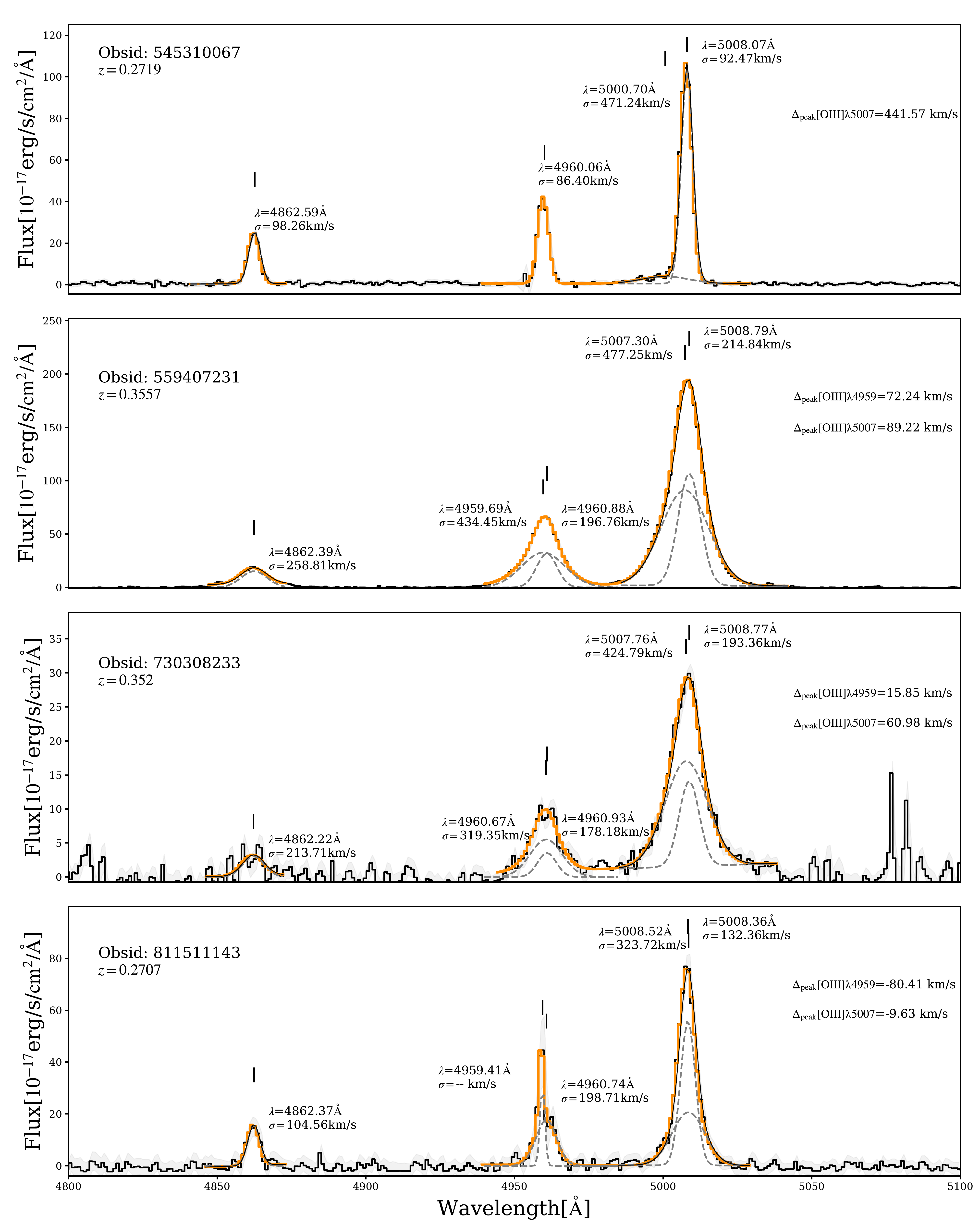}}
      \caption{Asymmetric [O~\textsc{iii}]$\lambda5007$ emission-line profiles of the first four sources.
      The black step plot is the re-calibrated LAMOST spectra with the continuum subtracted.
      The error region from the inverse variance is marked with the gray-shaded regions. 
      The central wavelengths from the spectral fit and the velocity dispersions of the profiles are marked above the emission lines.
      The velocity separation of the [O~\textsc{iii}]$\lambda5007$ emission line is marked on the right part of the plot.
      }
         \label{fig:asymmetry_1}
  \end{figure*}

\newpage
\begin{turnpage}

\begin{deluxetable*}{@{\extracolsep{6pt}}lllllllllll@{}}[!ht]
\tabletypesize{\scriptsize}
\tablecaption{Asymmetric [O~\textsc{iii}]$\lambda5007$ emission-line sources\label{tab:table_asymmetry}}
\tablehead{
\colhead{\textbf{obsid}} & \colhead{\textbf{SDSS objID}}  & \colhead{\textbf{$\rm\log[M_{\star}/M_{\odot}$]}}  & \colhead{\textbf{$\rm \log[\rm Age/year]$}} & \colhead{\textbf{z}}  & \colhead{\textbf{H$\beta$ }} &  \colhead{\textbf{[O~\textsc{iii}]$\lambda4959$}} & \colhead{\textbf{[O~\textsc{iii}]$\lambda5007$ }} & \colhead{\textbf{H$\alpha$ }} & \colhead{\textbf{[N \textsc{ii}]$\lambda6585$ }} 
 & \colhead{\textbf{BPT}}  
 \\
 \colhead{ } & \colhead{}  & \colhead{\textbf{}}  & \colhead{\textbf{}} & \colhead{\textbf{}}  & \colhead{\textbf{flux$^1$}} &  \colhead{\textbf{[flux}} & \colhead{\textbf{flux}} & \colhead{\textbf{flux}} & \colhead{\textbf{flux}} 
 & \colhead{\textbf{}}  
}
\startdata
173508168 & 12376787790009073330 & 10.46 & 7.22 & 0.3596 & 219.68 & 126.64 & 527.68 &1336.49 &  325.93 & comp \\
219710125 & 1237658492798304476 & 10.07 & 7.41 & 0.3627 & 237.18 & 98.29 & 457.23 & 860.11 &  337.10 & comp\\
285715023 & 1237664339319324770 & 9.33 & 8.96 & 0.1001 & 160.66 & 240.00 & 693.60 & 518.8 &  54.4 & -- \\
369415227 & 1237678847182701297 &  12.84& 7.00 & 0.3928 & 66.54 & 373.16 & 1206.71 & -- & --  & --\\
545310067 & 1237662620260761989 &  8.88 & 8.00 & 0.2719 & 122.26 & 189.29 & 518.23 & 181.32 & 9.73 & SF\\
559407231 & 1237658492798304476 & 10.07 & 7.41 & 0.3557 & 237.18 & 98.29 & 457.23 & 443.63 &  40.16 & comp\\
730308233 & 1237668350281777525 & 8.69 & 6.01 & 0.3520 & 47.50 & 105.87 & 352.50 & -- & -- &--\\
811511143 & 1237659132207104211 &  8.34 &6.40 & 0.2707&  81.22 &  218.78 &  634.63  & 395.44 & 197.77 & SF \\
\hline
\hline
\enddata
\tablenotetext{1}{The unit is $10^{-17}$ erg s$^{-1}$ cm$^{-2}$.}
\end{deluxetable*}
\clearpage
\end{turnpage}

\begin{deluxetable*}{@{\extracolsep{6pt}}cccccc@{}}[!ht]
\tabletypesize{\scriptsize}
\tablecaption{Single exposure measurement of the asymmetric [O~\textsc{iii}]$\lambda5007$ emission-line profiles \label{tab:single_exposure}}
\tablehead{
\colhead{\textbf{obsid}} & \colhead{\textbf{\# of exposures}} & \colhead{\textbf{\# of exposures used$^{2}$}} & 
\colhead{\textbf{$\bm \sigma[\lambda5007]$(km s$^{-1}$})} & \colhead{\textbf{$\bm \sigma[\lambda5007]$(wide)(km s$^{-1}$)}} & 
\colhead{\textbf{$\bm \Delta_{\rm peak}[\rm \lambda5007]$(km s$^{-1}$)}}   
}
\startdata
$173508168$ & $5$ &  $3$ & $163.11 \pm 3.12$ & $787.95 \pm 27.12$ & $310.12 \pm 150.43$ \\
$219710125$ & $6$ &  $3$ & $152.00 \pm 11.04$ & $510.73 \pm 141.61$ & $254.65 \pm 61.80$ \\
$285715023$ & $3$ &  $3$ & $48.95 \pm -- ^3$ & $442.77 \pm 67.31$ & $104.43 \pm 178.64$ \\
$369415227$ & $3$ &  $3$ & $186.40 \pm 8.14$ & $492.28 \pm 44.38$ & $114.56 \pm 62.17$ \\
$545310067$ & $2$ &  $2$ & $88.40 \pm 0.39$ & $453.93 \pm 35.33$ &  $488.63 \pm 69.43$ \\
$559407231$ & $3$ &  $3$ & $212.53 \pm 2.22$ & $483.04 \pm 5.45$ & $77.17 \pm 3.69$ \\
$730308233$ & $6$ &  $5$ & $202.88 \pm 49.79$ & $413.18 \pm 58.50$ & $94.05 \pm 30.71$ \\
$811511143$ & $4$ &  $3$ & $184.00\pm 6.09$ & $517.37\pm 9.51$ &  $24.56 \pm 21.61$  \\
\hline
\enddata
\tablenotetext{1}{The velocity dispersions in this table have been corrected from instrumental and thermal broadening.}
\tablenotetext{2}{We only use selected single exposures to calculate the velocity dispersion of the asymmetric [O~\textsc{iii}]$\lambda5007$ emission line, for the rest of the single exposures, the emission line is not detected.}
\tablenotetext{3}{Only one measurement is valid for the $\sigma[\lambda5007]$(km s$^{-1}$) measurement after subtracting the thermal broadening. }
\end{deluxetable*}

\begin{figure}[!ht]
  \centering
  \includegraphics[width=\hsize]{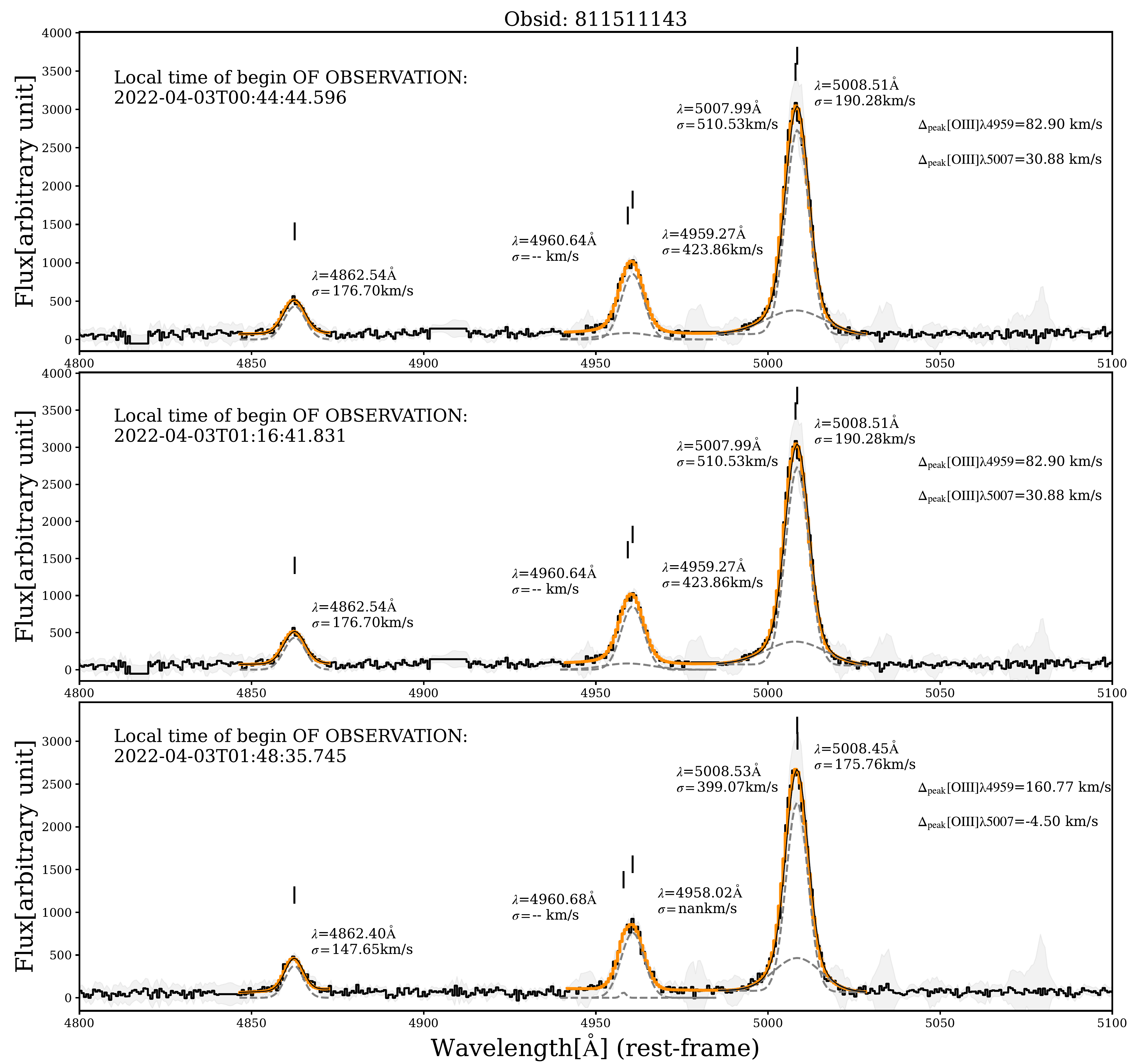}
      \caption{Asymmetric [O~\textsc{iii}]$\lambda4959$ and [O~\textsc{iii}]$\lambda5007$ emission-line profiles for each single exposure of 811511143.
      The black step plot is the original 1D spectra with the sky emission subtracted.
      The error region from the inverse variance is marked with the gray-shaded regions. 
      The central wavelengths from the spectral fit and the velocity dispersions of the profiles are marked above the emission lines.
      The velocity separations of the [O~\textsc{iii}]$\lambda4959$ and [O~\textsc{iii}]$\lambda5007$ emission lines are marked on the right part of the plot.
      }
         \label{fig:811511143}
  \end{figure}

\end{document}